\documentstyle[eqsecnum,aps,epsf]{revtex}
\newcommand{\cm}[1]{}

\newcommand{\tx}{\tilde{x}}

\newcommand{\trho}{\tilde{\rho}}
\newcommand{\mean}[1]{\left\langle\,#1\,\right\rangle^{\Omega,x_{m}}_{x_b,x_a}}
\newcommand{\meandA}[1]{\left\langle\,#1\,\right\rangle^{\Omega,x_{m}}_{x_a,x_a}}
\newcommand{\meand}[1]{\left\langle\,#1\,\right\rangle^{\Omega}_{x_a,x_a}}
\newcommand{\meaniso}[1]{\left\langle\,#1\,\right\rangle^\Omega_{{\bf r}_a,{\bf r}_a}}
\newcommand{\meananiso}[1]{\left\langle\,#1\,\right\rangle^{\Omega_{L,T}}_{{\bf r}_a,{\bf r}_a}}
\newcommand{\cum}[1]{\left\langle\,#1\,\right\rangle^{\Omega,x_{m}}_{x_b,x_a,c}}
\newcommand{\cumd}[1]{\left\langle\,#1\,\right\rangle^{\Omega}_{x_a,x_a,c}}
\newcommand{\blambda}{\mbox{\boldmath$\lambda$}}

\begin{document}
\title{Variational Perturbation Theory for Density Matrices}
\author{M. Bachmann, H. Kleinert, and A. Pelster}
\address{Institut f\"ur Theoretische Physik, Freie Universit\"at Berlin, Arnimallee 14, 14195 Berlin}
\date{\today}
\maketitle
\begin{abstract}
We develop convergent variational perturbation theory
for quantum statistical density matrices.
The theory is applicable
to polynomial as well as nonpolynomial
interactions.
Illustrating the power of the theory,
 we calculate the temperature-dependent density of a particle
in a double-well
and of the electron in a hydrogen atom.
\end{abstract}
\section{Introduction}
Variational perturbation theory~\cite{kl213,PI}
transforms divergent perturbation expansions into convergent ones.
The convergence extends to infinitely strong couplings~\cite{conv},
a property which has recently been
used to derive critical exponents in field theory
without renormalization group methods~\cite{kl257,kl263}.
The theory has first been
developed in quantum mechanics for the path integral representation
of the free energy
of the anharmonic oscillator~\cite{kl220}
and the hydrogen atom~\cite{PI,kl267}.
Local quantities such as quantum statistical density matrices
have been
treated so far only to lowest-order for the anharmonic oscillator
and the hydrogen atom~\cite{kl145,kl153}.
There has also been a related first-order treatment in
classical phase space~\cite{cucc1} for systems with
dissipation~\cite{cucc2}.

The purpose of this paper is to develop a
systematic convergent variational perturbation theory for
the path integral representation
of density matrices
of a point particle moving in a polynomial or nonpolynomial
potential.
As a first application we calculate the particle density in a double-well
and then the electron
density in a hydrogen atom.
\section{General Features}
\label{genfeat}
Variational perturbation theory approximates a
 quantum statistical system
by perturbation expansions around harmonic oscillators
with
trial frequencies which are optimized
differently for each order of the expansions.
When dealing with the free energy,
it is essential to give a special treatment to the
fluctuations of the path average $\overline x\equiv  (k_B T/\hbar)\int _0^{\hbar/k_BT }d\tau \,x(\tau )$,
since this performs violent fluctuations at high
temperatures $T$. These cannot be treated by any expansion,
unless the potential is close to harmonic.
The effect of these fluctuations may, however, easily be calculated at the end
by a single numerical fluctuation integral.
For this reason, variational perturbation expansions
are performed for each position $x_0$ of the path average
separately, yielding an $N$th order approximation
$W_N(x_0)$ to the local free energy $V_{\rm eff,cl}(x_0)$, called the
{\em effective classical potential\/}\cite{effcl}.
The name indicates that one may obtain the full quantum partition function
$Z$
from this object by a simple integral over $x_0$
just as in classical statistics,
\begin{equation}
  \label{zmpfeff}
  Z=\int_{-\infty}^{+\infty}\frac{dx_0}{\sqrt{2\pi\hbar^2/Mk_BT}}\,\exp\left\{-V_{\rm eff,cl}(x_0)/k_BT\right\}.
\end{equation}
Having calculated $W_N(x_0)$, we obtain the $N$th-order approximation to the partition function
\begin{equation}
Z_N=\int\limits_{-\infty}^{+\infty}\frac{dx_0}{\sqrt{2\pi\hbar^2/M k_B T}} e^{-
W_N(x_0)/k_BT}.
\label{gf1}
\end{equation}
The separate treatment
of the path average is important to ensure a
fast convergence at larger temperatures. In the high-temperature limit,
$W_N(x_0)$
converges against the initial potential $V(x_0)$ for any order $N$.

Before embarking upon the theory, it is useful to
visualize some characteristic properties of path fluctuations.
Consider the euclidean path integral
over all periodic paths
$x(\tau )$, with $x(0)=x(\hbar/k_BT)$,
for a harmonic oscillator with minimum at
$x_m$, where the action is
\begin{equation}
  \label{zmharmaction}
  {\cal A}_{\Omega,x_m}[x]=\int_0^{\hbar/k_BT}d\tau\,\left\{\frac{1}{2}M\dot{x}^2(\tau)+\frac{1}{2}M\Omega^2[x(\tau)-x_m]^2 \right\}.
\end{equation}
Its partition function is
\begin{equation}
  \label{zmpf}
  Z^{\Omega,x_m}=\oint {\cal D}x\,\exp\left\{-{\cal A}_{\Omega,x_m}[x]/\hbar \right\}=\frac{1}{2\sinh\hbar\Omega/2k_BT}
\end{equation}
and the correlation functions
of local quantities $O_1(x)$, $O_2(x)$, \dots are given by the expectation values
\begin{equation}
  \label{meanval}
  \langle\,
O_1(x(\tau))
O_2(x(\tau))\cdots\,
\,\rangle^{\Omega,x_m}=\frac{1}{Z^{\Omega,x_m}}\oint {\cal D}x\,O_1(x(\tau_1))O_2(x(\tau_2))\,\cdots \exp\left\{-{\cal A}_{\Omega,x_m}[x]/\hbar \right\}.
\end{equation}
The particle distribution of the oscillator is given by
\begin{equation}
  \label{zmdfQS}
  P_H(x)\equiv\langle\,\delta(x-x(\tau))\,\rangle^{\Omega,x_m}
=\frac{1}{\sqrt{2\pi a^2_{\rm H}}}\exp\left[-\frac{(x-x_m)^2}{2 a_{\rm H}^2} \right],
\end{equation}
which is a Gaussian distribution
of width
\begin{equation}
  \label{zmwidQS}
  a_{\rm H}^2=\frac{\hbar}{2M\Omega}\coth\frac{\hbar\Omega}{2 k_BT},
\end{equation}
the subscript
indicating that we are dealing with a harmonic oscillator.
At zero temperature,
this is equal to the square of the ground-state wave function
of the harmonic oscillator, whose width is
\begin{equation}
  \label{qmflwidth}
  a^2_{\rm H\,0}=\frac{\hbar}{2M\Omega}.
\end{equation}
\begin{figure}[t]
\centerline{
\setlength{\unitlength}{1cm}
\begin{picture}(12.0,9.0)
\put(0,0){\makebox(12,9){\epsfxsize=12cm \epsfbox{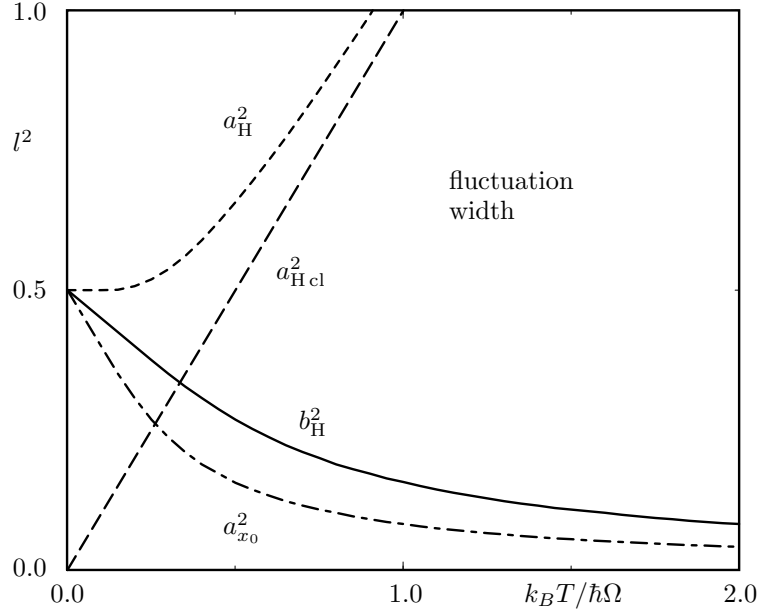}}}
\put(5,3){$b_{\rm H}^2$}
\put(4,7){$a_{\rm H}^2$}
\put(4,1.6){$a_{x_0}^2$}
\put(4.7,5){$a_{\rm H\,cl}^2$}
\put(8,0.7){$ k_B T/ \hbar\Omega $}
\put(1.2,6.7){$l^2$}
\put(7,6.2){fluctuation}
\put(7,5.8){width}
\put(1.7,0.7){0.0}
\put(6.2,0.7){1.0}
\put(10.65,0.7){2.0}
\put(1.2,1.05){0.0}
\put(1.2,4.75){0.5}
\put(1.2,8.45){1.0}
\end{picture}
}
\caption{\label{wid}
 Temperature dependence of fluctuation widths
of any point $x(\tau)$ on the path in a harmonic oscillator
($l^2$ is a square length in units of $\hbar/M\Omega$).
The quantity $a^2_{\rm H}$ (dashed) is the quantum mechanical width, whereas $a_{x_0}^2$ (dash-dotted) shares the width after separating out the fluctuations around the path average $x_0$. The quantity $a_{\rm H\,cl}^2$ (long-dashed) is the width of the classical distribution, and
 $b^2_{\rm H}$ (solid curve) is the fluctuation width at fixed ends which is relevant for the calculation of the density matrix by variational perturbation theory.}
\end{figure}
In the limit $\hbar\to 0$, we obtain from (\ref{meanval}), (\ref{zmdfQS}) the classical distribution
\begin{equation}
  \label{zmdfcl}
  P_{\rm H\,cl}(x)=\frac{1}{\sqrt{2\pi a^2_{\rm H\,cl}}}\exp\left[-\frac{(x-x_m)^2}{2 a^2_{\rm H\,cl}} \right]
\end{equation}
with
\begin{equation}
  \label{zmwidcl}
  a^2_{\rm H\,cl}= \frac{k_BT}{M\Omega^2}.
\end{equation}
The linear growth of this classical width is the origin of the
famous Dulong-Petit law
for the specific heat of a harmonic system.
The classical fluctuations are governed by the integral over the Boltzmann factor
\begin{equation}
  \label{boltzmann}
  e^{-M\Omega^2(x-x_m)^2/2k_BT},
\end{equation}
in the classical partition function
\begin{equation}
Z_{\rm H\,cl}=\int \limits_{-\infty}^{+\infty}\frac{dx}{\sqrt{2\pi\hbar^2/M k_B T}} e^{-M\Omega^2(x-x_m)^2/2k_B T}.
\label{gf2}
\end{equation}
From this we obtain the classical distribution (\ref{zmdfcl}) as the expectation value
\begin{equation}
  \label{gf3}
  P_{\rm H\,cl}(x)\equiv \langle\,\delta(x-\overline x)\,\rangle^{\Omega,x_m}_{\rm cl}=Z_{\rm cl}^{-1}\int\limits_{-\infty}^{+\infty} \frac{d\overline x}{\sqrt{2\pi\hbar^2/M k_B T}}\, \delta (x-\overline x)\,e^{-M\Omega^2(\overline x-x_m)^2/2k_B T}
=
\frac{1}{\sqrt{2\pi a^2_{\rm cl}}}\exp\left[-\frac{(x-x_m)^2}{2 a^2_{\rm cl}} \right].
\end{equation}
Variational perturbation theory avoids the divergence of the harmonic width $a_{\rm H}^2$ at high temperatures (\ref{zmwidcl}) by
the separate
treatment of the fluctuations of the path average $\overline x$, as explained above.
The average is fixed at some value $x_0$
with the help of a delta function $\delta(\overline{x}-x_0)$.
For each $x_0$ we introduce local expectation values
\begin{equation}
  \label{zmcffk}
  \langle\,O_1(x(\tau_1))O_2(x(\tau_2))\cdots\,\rangle^{\Omega,x_m}_{x_0}=\frac{\langle\,\delta(\overline{x}-x_0)\,O_1(x(\tau_1))\,O_2(x(\tau_2))\cdots \,\rangle^{\Omega,x_m}}{\langle\,\delta(\overline{x}-x_0) \,\rangle^{\Omega,x_m}}.
\end{equation}
The original quantum statistical distribution
of the harmonic oscillator
(\ref{zmdfQS})
collects fluctuations of $\overline x=x_0$ and those around $x_0$, and can therefore written as a convolution
\begin{equation}
  \label{zmconv}
  P_{\rm H}(x)=\int_{-\infty}^{+\infty}dx_0\,P_{x_0}(x-x_0)\,P_{\rm cl}(x_0),
\end{equation}
over  the classical distribution (\ref{zmdfcl}) and the
local one:
\begin{equation}
  \label{zmdffk}
P_{x_0}(x)=\langle\,\delta(x-x(\tau))\,\rangle^{\Omega,x_m}_{x_0}=\frac{1}{\sqrt{2\pi a_{x_0}^2}}\,\exp\left[-\frac{(x-x_0)^2}{2a_{x_0}^2} \right].
\end{equation}
Such a convolution of Gaussian distributions (\ref{zmconv}) leads to
another Gaussian distribution with added
widths, so that the width of the local
distribution is
given by the difference
\begin{equation}
  \label{zmwidfk}
  a^2_{x_0}=a^2_{\rm H}-a^2_{\rm cl}=
\frac{\hbar}{2M\Omega}\left(\coth\frac{\hbar\Omega}{2k_BT}-\frac{2k_BT}{\hbar\Omega} \right),
\end{equation}
which starts out at a finite value for $T=0$, and goes to zero for $T\to\infty$,
\begin{equation}
  \label{wqmlim}
  \lim_{T\to\infty}a^2_{x_0}=\frac{\hbar\Omega}{12 k_BT}.
\end{equation}
The latter property suppresses all fluctuations around $\overline x$ and guarantees that $\lim_{T\to\infty}
W_N(x_0)=V(x_0)$ for all $N$ (see Fig. \ref{wid}).

With this separation of the path average,
the partition function
\begin{equation}
  \label{zmgenpf}
  Z=\oint{\cal D}x\,\exp\left\{-{\cal A}[x]/\hbar \right\}
\end{equation}
 for the general particle action
\begin{equation}
  \label{zmgenact}
  {\cal A}[x]=\int_0^{\hbar/k_BT}d\tau\,\left[\frac{1}{2}M\dot{x}^2(\tau)+V(x(\tau)) \right]
\end{equation}
possesses the effective classical representation (\ref{zmpfeff}) with the
effective classical potential
\begin{equation}
  \label{zmeffpot}
  V_{\rm eff,cl}(x_0) = -k_BT\,\ln\left(\sqrt{\frac{2\pi\hbar^2}{Mk_BT}}\oint{\cal D}x\,\delta(x_0-\overline{x})\,\exp\left\{-{\cal A}[x]/\hbar \right\} \right).
\end{equation}
In variational perturbation theory, this is expanded perturbatively around an $x_0$-dependent harmonic system with trial frequency $\Omega(x_0)$, whose optimization leads to the approximation $
W_N(x_0)$ for $V_{\rm eff,cl}(x_0)$.
\section{Density Matrix of Harmonic Oscillator}
How can this method be extended to density matrices?
Their path integral representation is
\begin{equation}
  \label{zmdm}
  \rho(x_b,x_a)=\frac{1}{Z}\trho(x_b,x_a)
\end{equation}
where $\trho(x_b,x_a)$ is the path integral
\begin{equation}
\label{zmprop}
  \trho(x_b,x_a)=\int\limits_{(x_a,0)\leadsto (x_b,\hbar/k_BT)}{\cal D}x\,\exp\left\{-{\cal A}[x]/\hbar \right\}
\end{equation}
over all paths with the fixed endpoints $x(0)=x_a$ and $x(\hbar/k_BT)=x_b$.
The partition function is found from the trace of $\trho(x_b,x_a)$:
\begin{equation}
  \label{pfi}
  Z = \int\limits_{-\infty}^{+\infty}dx\,\trho(x,x).
\end{equation}
For a harmonic oscillator centered at $x_m$ (\ref{zmharmaction}), the path integral (\ref{zmprop}) can be easily done with the result~\cite{PI}
\begin{equation}
  \label{harmdens}
  \trho_0^{~\Omega,x_m}(x_b,x_a)=\sqrt{\frac{M\Omega}{2\pi\hbar \sinh\hbar\Omega/k_BT}}\exp\left\{-\frac{M\Omega}{2\hbar \sinh{\hbar\Omega/k_BT}}\left[(\tx_b^2+\tx_a^2)\cosh{\hbar\Omega/k_BT}-2 \tx_b \tx_a\right]\right\}
\end{equation}
where
\begin{equation}
  \label{displ}
  \tx(\tau)=x(\tau)-x_m.
\end{equation}
At fixed endpoints $x_b,x_a$, the quantum mechanical correlation functions are
\begin{equation}
  \label{zmdmmv}
  \mean{O_1(x(\tau_1))\,O_2(x(\tau_2))\cdots}=\frac{1}{
\trho_0^{~\Omega,x_m}(x_b, x_a)}\,\int\limits_{(x_a,0)\leadsto (x_b,\hbar/k_BT)}{\cal D}x\,O_1(x(\tau_1))\,O_2(x(\tau_2))\cdots\,\exp\left\{-{\cal A}_{\Omega,x_m}[x]/\hbar \right\}
\end{equation}
and the distribution function is given by
\begin{equation}
  \label{zmdmQS}
  p_{\rm H}(x,\tau)\equiv \mean{\delta(x-x(\tau))} = \frac{1}{\sqrt{2\pi b^2_{\rm H}(\tau)}}\exp\left[-\frac{(\tx-x_{\rm cl})^2}{2 b^2_{\rm H}(\tau)} \right].
\end{equation}
The classical path of a particle in a harmonic potential is
\begin{equation}
  \label{zmclpath}
  x_{\rm cl}(\tau) = \frac{\tx_b\sinh \Omega\tau+\tx_a\sinh\Omega(\hbar/k_BT-\tau) }{{\sinh \hbar\Omega/k_BT}}
\end{equation}
and the time-dependent width $b^2_{\rm H}(\tau)$ is found to be
\begin{equation}
  \label{zmdmQSwid}
  b^2_{\rm H}(\tau)=\frac{\hbar}{2M\Omega}\left\{\coth \frac{\hbar\Omega}{k_BT} -\frac{\cosh[\Omega(2\tau-\hbar/k_BT)]}{\sinh \hbar\Omega/k_BT}\right\}.
\end{equation}
Since the euclidean time $\tau$ lies in the interval $0\le\tau\le \hbar/k_BT$, the width (\ref{zmdmQSwid}) is bounded by
\begin{equation}
  b_{\rm H}^2(\tau)\le \frac{\hbar}{2M\Omega}\tanh\frac{\hbar\Omega}{2k_BT},
\end{equation}
thus remaining finite at all temperatures. The temporal average of (\ref{zmdmQSwid}) is
\begin{equation}
  \label{zmdmQSwidav}
  b^2_{\rm H}=\frac{k_BT}{\hbar}\int_0^{\hbar/k_BT}d\tau\,b^2_{\rm H}(\tau)=\frac{\hbar}{2M\Omega}\left(\coth\frac{\hbar\Omega}{k_BT}-\frac{k_BT}{\hbar\Omega}\right).
\end{equation}
Just as $a_{x_0}^2$, this goes to zero for $T\to\infty$ with an asymptotic behaviour $\hbar\Omega/6k_BT$, which is twice as big as that of $a_{x_0}^2$ (see Fig.~\ref{wid}).
\section{Variational Perturbation Theory for Density Matrices}
\label{theory}
To obtain a variational approximation for the density matrix,
it is useful to separate the general action (\ref{zmgenact})
into a trial one for which the euclidean propagator is known,
and a remainder containing the original potential.
If we were to  proceed in complete analogy with the treatment
 of the partition function,
we would expand the euclidean path integral around
a trial harmonic one with fixed end points $x_b,x_a$ and
a fixed path average $x_0$, and with a trial frequency $\Omega(x_b,x_a;x_0)$. The result would be
an
effective classical potential $
W_N(x_b,x_a;x_0)$ to be optimized in $\Omega(x_b,x_a;x_0)$.
After that we would have to perform a final integral in $x_0$ over the Boltzmann factor $\exp[-
W_N(x_b,x_a;x_0)/k_BT]$.
\begin{figure}
\centerline{
\setlength{\unitlength}{1cm}
\begin{picture}(12.0,9.0)
\put(0,0){\makebox(12,9){\epsfxsize=12cm \epsfbox{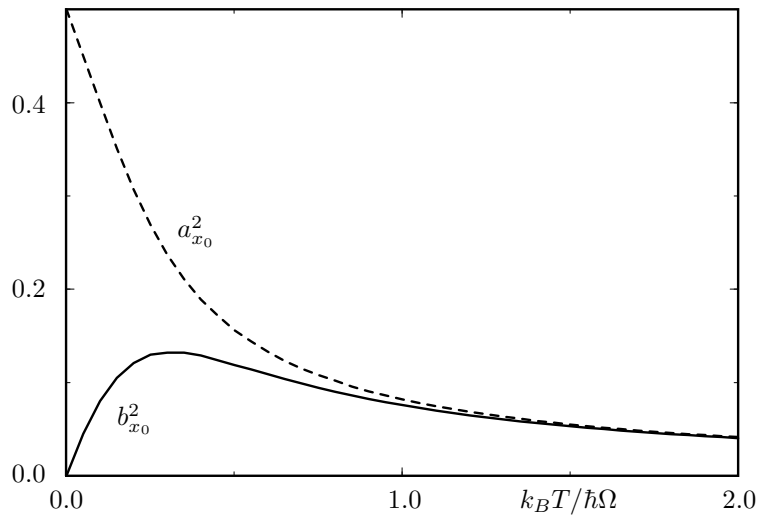}}}
\put(1.7,0.7){0.0}
\put(6.2,0.7){1.0}
\put(7.95,0.7){$k_BT/\hbar\Omega$}
\put(10.65,0.7){2.0}
\put(1.2,1.05){0.0}
\put(1.2,3.5){0.2}
\put(1.2,5.95){0.4}
\put(3.4,4.3){$a_{x_0}^2$}
\put(2.6,1.8){$b_{x_0}^2$}
\end{picture}
}
\caption{\label{widB}
Temperature-dependence of the width of fluctuations around the path average $x_0=\overline x$ at fixed ends. For comparison we also show the width $a_{x_0}^2$ of
Fig.~\protect\ref{wid}. The vertical axis gives these square lengths
in units of $\hbar/M\Omega$ again.}
\end{figure}
But, because of the finiteness of the fluctuation width $b_{\rm H}^2$ at all temperatures which is similar to that of $a_{x_0}^2$, the special treatment of $\overline x=x_0$ becomes superfluous for paths with fixed endpoints $x_b,x_a$.
While the separation of $x_0$ was necessary to deal with the diverging fluctuation width of the path average $\overline x$, paths with fixed ends have fluctuations of the path average which are governed by the distribution
\begin{equation}
  \label{vpt1}
  p_{x_0}(x_b,x_a;x_0)\equiv\mean{\delta(x-\overline x)}=\frac{1}{\sqrt{2\pi b^2_{x_0}}}\exp\left\{-\frac{1}{2b_{x_0}^2}\left[\tx_0-\frac{1}{2}(\tx_b+\tx_a)\frac{2k_BT}{\hbar\Omega}\tanh{\frac{\hbar\Omega}{2 k_BT}} \right]^2\right\}
\end{equation}
with the width
\begin{equation}
  \label{vpt2}
b_{x_0}^2=\frac{k_BT}{M\Omega^2}\left[1-\frac{2k_BT}{\hbar\Omega}\tanh{\frac{\hbar\Omega}{2k_BT}} \right],
\end{equation}
which goes to zero for both limits $T\to 0$ and $T\to\infty$ (see Fig.~\ref{widB}).
At each euclidean time, $x(\tau)$ fluctuates narrowly around the classical path $x_{\rm cl}(\tau)$ connecting $x_b$ and $x_a$. This is the reason why we may treat the fluctuations of $\overline x=x_0$ by variational perturbation theory, just as the other fluctuations. As a remnant of the extra treatment of $x_0$ we must, however, perform the initial perturbation expansion around the
minimum of the
effective classical potential which will lie at some point $x_m$ determined by the endpoints $x_b,x_a$, and by the minimum of the potential $V(x)$. Thus we shall use the euclidean path integral for the density matrix of the harmonic oscillator centered at $x_m$ as the trial system around which to perform the variational perturbation theory, treating the fluctuations of $x_0$ around $x_m$ on the same footing as the remaining fluctuations. The position $x_m$ of the minimum is a function $x_m=x_m(x_b,x_a)$, and has to be optimized with respect to the trial frequency, which itself is a function $\Omega=\Omega(x_b,x_a)$
 to be optimized.

Hence we start by decomposing the action (\ref{zmgenact}) as
\begin{equation}
  \label{actsep}
  {\cal A}[x]={\cal A}_{\Omega,x_m}[x]+{\cal A}_{\rm int}[x]
\end{equation}
with an interaction
\begin{equation}
  \label{acti}
{\cal A}_{\rm int}[x(\tau)]=\int_0^{\hbar \beta}d\tau\,V_{\rm int}(x(\tau)),
\end{equation}
where the interaction potential is the difference between the original one $V(x)$ and the inserted displaced harmonic oscillator:
\begin{equation}
  \label{potint}
  V_{\rm int}(x(\tau))=V(x(\tau))-\frac{1}{2}M\Omega^2[x(\tau)-x_m]^2.
\end{equation}
For brevity, we have introduced the inverse temperature
in natural units
$\beta\equiv 1/k_BT$ in (\ref{acti}).
Now we evaluate the path integral for the euclidean propagator (\ref{zmprop})
by treating the interaction (\ref{acti}) as a perturbation, leading to a moment expansion
\begin{equation}
  \label{momexp}
  \trho(x_b,x_a)=\trho_0^{~\Omega,x_m}(x_b,x_a)\left[1-\frac{1}{\hbar}\mean{{\cal A}_{\rm int}[x]}+\frac{1}{2\hbar^2}\mean{{\cal A}_{\rm int}^2[x]}-\ldots  \right],
\end{equation}
with expectation values defined in (\ref{zmdmmv}). The zeroth order consists of the harmonic contribution (\ref{harmdens}) and higher orders contain harmonic averages of the interaction (\ref{acti}). The correlation functions in (\ref{momexp}) can be decomposed into connected ones by going over to cumulants, yielding
\begin{equation}
\label{nd}
\trho(x_b,x_a)=\trho_0^{~\Omega,x_m}(x_b,x_a)\exp\left[-\frac{1}{\hbar}\cum{{\cal A}_{\rm int}[x]}+\frac{1}{2\hbar^2}\cum{{\cal A}_{\rm int}^2[x]}-\ldots\right],
\end{equation}
where the first cumulants are defined as usual
\begin{eqnarray}
  \label{cums}
  \cum{O_1(x(\tau_1))}&=&\mean{O_1(x(\tau_1))},\nonumber\\
  \cum{O_1(x(\tau_1))O_2(x(\tau_2))}&=&\mean{O_1(x(\tau_1))O_2(x(\tau_2))}-\mean{O_1(x(\tau_1))}\mean{O_2(x(\tau_2))},\nonumber\\
&\vdots&\quad .
\end{eqnarray}
The series (\ref{nd}) is truncated after the $N$-th term, resulting in the $N$-th order approximant for the quantum statistical density matrix
\begin{equation}
  \label{ndx}
  \trho_N^{~\Omega,x_m}(x_b,x_a)=\trho_0^{~\Omega,x_m}(x_b,x_a)\exp\left[\sum\limits_{n=1}^N\,\frac{(-1)^n}{n! \hbar^n}\,\cum{{\cal A}_{\rm int}^n[x]}\right],
\end{equation}
which explicitly depends on both variational parameters $\Omega$ and $x_m$.

In analogy
to classical statistics, where the Boltzmann distribution
in configuration space is controlled
by the classical potential $V(x)$ according to
\begin{equation}
  \label{dcl}
  \trho_{\rm cl}(x)=\sqrt{\frac{M}{2\pi\hbar^2\beta}}\exp\left[-\beta V(x) \right],
\end{equation}
we now
introduce a new type of
{\em effective classical potential\/}
 $V_{\rm eff,cl}(x_a,x_b)$ which governs
the unnormalized density matrix
\begin{equation}
  \label{deff}
  \trho(x_b,x_a)=\sqrt{\frac{M}{2\pi\hbar^2\beta}}\exp\left[-\beta V_{\rm eff,cl}(x_b,x_a) \right].
\end{equation}
Its $N$th order approximation is
obtained from (\ref{harmdens}), (\ref{ndx}), and (\ref{deff})
voa the cumulant expansion
\begin{equation}
  \label{veff}
  W_N^{\Omega,x_m}(x_b,x_a)=\frac{1}{2\beta}\ln{\frac{\sinh{\hbar\beta\Omega}}{\hbar\beta\Omega}}+\frac{M\Omega}{2\hbar\beta\sinh{\hbar\beta\Omega}}\left\{(\tx_b^2+\tx_a^2)\cosh{\hbar\beta\Omega}-2\tx_b\tx_a\right\}-\frac{1}{\beta}\sum\limits_{n=1}^N\,\frac{(-1)^n}{n!\hbar^n}\,\cum{{\cal A}_{\rm int}^n[x]},
\end{equation}
which is  optimized for each set of endpoints $x_b,x_a$
in the variational parameters $\Omega^2$ and $x_m$,
the result being denoted by
$W_N(x_b,x_a)$.
The optimal values $\Omega^2(x_a,x_b)$ and $x_m(x_a,x_b)$ are determined from the
extremality conditions\begin{equation}
  \label{mincond}
  \frac{\partial
W_N^{\Omega,x_m}(x_b,x_a)}
{\partial \Omega^2}\stackrel{!}{=}0,\quad\frac{\partial
W_N^{~\Omega,x_m}(x_b,x_a)}{\partial x_m}\stackrel{!}{=}0.
\end{equation}
The solutions are
denoted by  ${\Omega^2}^{N},x_m^N $, both being functions
of $x_b,\,x_a$.
 If no extrema are found, one has to look for the flattest region of the function (\ref{veff}), where the lowest higher-order derivative disappears. Eventually the normalized density matrix is obtained from
\begin{equation}
  \label{normdx}
  \rho_N(x_b,x_a)=Z_N^{-1}
\trho_N^{{~\Omega^2}^{N},x_m^N }(x_b,x_a ),
\end{equation}
where
\begin{equation}
  \label{vptpf}
  Z_N=\int_{-\infty}^{+\infty}dx\,
\trho_N^{~{\Omega^2}^{N},x_m^N }(x_b,x_a ),
\end{equation}
In principle, one could also optimize the entire ratio (\ref{normdx}), but this would be harder to do in practice. Moreover, the optimization of the unnormalized density matrix is the only option, if the normalization diverges due to singularities of the potential.
This will be seen in Sect.~\ref{csect}.
\section{Smearing Formula for Density Matrices}
\label{smform}
In order to calculate the connected correlation functions in the variational perturbation expansion (\ref{ndx}), we must find efficient formulas for evaluating expectation values (\ref{zmdmmv}) of any power of the interaction (\ref{acti})
\begin{equation}
  \label{nmean}
  \mean{{\cal A}^n_{\rm int}[x]}=\frac{1}{\trho_0^{~\Omega,x_m}(x_b,x_a)}\int\limits_{\tx_a,0}^{\tx_b,\hbar\beta}{\cal D}\tx\,\prod\limits_{l=1}^n\left[\int_0^{\hbar\beta}d\tau_l\,V_{\rm int}(\tx(\tau_l)+x_m)\right]\,\exp\left\{-\frac{1}{\hbar}{\cal A}_{\Omega,x_m}[\tx+x_m] \right\}.
\end{equation}
This can be done by an extension of the smearing formalism, developed in Ref.~\cite{kl267}.
We rewrite the interaction potentials as
\begin{equation}
  V_{\rm int}(\tx(\tau_l)+x_m)=\int\limits_{-\infty}^{+\infty}dz_l\,V_{\rm int}(z_l+x_m)\int\limits_{-\infty}^{+\infty}\frac{d\lambda_l}{2\pi}\,\exp\{i\lambda_lz_l\}\,\exp\left[-\int_0^{\hbar\beta}d\tau\,i\lambda_l\delta(\tau-\tau_l)\tx(\tau) \right]
\end{equation}
and introduce a current
\begin{equation}
  \label{jall}
  J(\tau)=\sum\limits_{l=1}^n\,i\hbar\lambda_l\delta(\tau-\tau_l),
\end{equation}
so that (\ref{nmean}) becomes
\begin{equation}
\label{nmeanB}
\mean{{\cal A}^n_{\rm int}[x]}=\frac{1}{\trho_0^{~\Omega,x_m}(x_b,x_a)}\prod\limits_{l=1}^n\left[\int_0^{\hbar\beta}d\tau_l\,\int_{-\infty}^{+\infty}dz_l\,V_{\rm int}(z_l+x_{\rm min})\,\int_{-\infty}^{+\infty}\frac{d\lambda_l}{2\pi}\,\exp\{i\lambda_l z_l \}\right]\,
K^{\Omega,x_m}[J].
\end{equation}
The kernel $K^{\Omega,x_m}[J]$ represents the generating functional for all correlation functions of the displaced harmonic oscillator
\begin{equation}
  \label{kernel}
  K^{\Omega,x_m}[J]=\int\limits_{\tx_a,0}^{\tx_b,\hbar\beta}{\cal D}\tx\,\exp\left\{-\frac{1}{\hbar}\int_0^{\hbar\beta}d\tau\,\left[\frac{m}{2}\dot{\tx}^2(\tau)+\frac{1}{2}M\Omega^2\tx^2(\tau)+J(\tau)\,\tx(\tau) \right]\right\}.
\end{equation}
For zero current $J$, this generating functional reduces the euclidean harmonic propagator (\ref{harmdens}):
\begin{equation}
  K^{\Omega,x_m}[J=0]=\trho_0^{~\Omega,x_m}(x_b,x_a).
\end{equation}
For nonzero $J$, the solution of the functional integral (\ref{kernel}) is given by
\begin{equation}
  \label{solkern}
  K^{\Omega,x_m}[J]=\trho_0^{~\Omega,x_m}(x_b,x_a)\exp\left[-\frac{1}{\hbar}\int_0^{\hbar\beta}d\tau\,J(\tau)\,x_{\rm cl}(\tau)+\frac{1}{2\hbar^2}\int_0^{\hbar\beta}d\tau\,\int_0^{\hbar\beta}d\tau'\,J(\tau)\,
G^\Omega(\tau,\tau')\,J(\tau') \right],
\end{equation}
where $x_{\rm cl}(\tau)$ denotes the classical path (\ref{zmclpath}) and $G^\Omega(\tau,\tau')$ the harmonic Green function
\begin{equation}
  \label{ggreenh}
  G^\Omega(\tau,\tau')=\frac{\hbar}{2M\Omega}\frac{\cosh{\Omega(|\tau-\tau'|-\hbar\beta)}-\cosh{\Omega(\tau+\tau'-\hbar\beta)}}{\sinh{\hbar\beta\Omega}}.
\end{equation}
The expression (\ref{solkern}) can be simplified by using the explicit expression (\ref{jall}) for the current $J$. This leads to a generating functional
\begin{equation}
  \label{solkernB}
  K^{\Omega,x_m}[J]=\trho_0^{~\Omega,x_m}(x_b,x_a)\,\exp\left(-i\blambda^T{\bf x}_{\rm cl} -\frac{1}{2}\,\blambda^T\,G\,\blambda \right),
\end{equation}
where we have introduced the $n$--dimensional vectors $\blambda=(\lambda_1,\ldots,\lambda_n)^T$, ${\bf x}_{\rm cl}=(x_{\rm cl}(\tau_1),\ldots,x_{\rm cl}(\tau_n))^T$
with the superscript $T$ denoting transposition, and the symmetric $n\times n$-matrix $G$ whose elements are
$G_{kl}=G^\Omega(\tau_k,\tau_l)$. Inserting (\ref{solkernB}) into (\ref{nmeanB}), and performing the integrals with respect to $\lambda_1,\ldots,\lambda_n$, we obtain the $n$-th order smearing formula for the density matrix
\begin{eqnarray}
  \label{smear}
  \mean{{\cal A}_{\rm int}^n[x]}&=&\prod\limits_{l=1}^n\left[\int_0^{\hbar\beta}d\tau_l\int_{-\infty}^{+\infty}dz_l\,V_{\rm int}(z_l+x_m)\right]\nonumber\\
& &\times\frac{1}{\sqrt{(2\pi)^n\,{\rm det}\,G}}\,\exp\left\{-\frac{1}{2}\sum\limits_{k,l=1}^n\,[z_k-x_{\rm cl}(\tau_k)]\,G^{-1}_{kl}\,[z_l-x_{\rm cl}(\tau_l)] \right\}.
\end{eqnarray}
The integrand contains an $n$-dimensional Gaussian distribution describing both thermal and quantum fluctuations around the harmonic classical path $x_{\rm cl}(\tau)$ of Eq. (\ref{zmclpath}) in a trial oscillator centered at $x_m$, whose width is governed by the Green functions (\ref{ggreenh}).

For closed paths with coinciding endpoints ($x_b=x_a$), formula (\ref{smear}) leads to the $n$-th order smearing formula for particle densities
\begin{equation}
  \label{normd}
  \rho(x_a)=\frac{1}{Z}\trho(x_a,x_a)=\frac{1}{Z}\oint{\cal D}x\,\delta(x(\tau=0)-x_a)\,\exp\{-{\cal A}[x]/\hbar\},
\end{equation}
which can be written as
\begin{equation}
  \label{smeard}
  \meandA{{\cal A}_{\rm int}^n[x]}=\frac{1}{\rho_0^{\Omega,x_m}(x_a)}\prod\limits_{l=1}^n\left[\int_0^{\hbar\beta}d\tau_l\int_{-\infty}^{+\infty}dz_l\,V_{\rm int}(z_l+x_m)\right]\frac{1}{\sqrt{(2\pi)^{n+1}\,{\rm det}\,a^2}}\,\exp\left(-\frac{1}{2}\sum\limits_{k,l=0}^n\,z_k\,a^{-2}_{kl}\,z_l \right)
\end{equation}
with $z_0=\tx_a$. Here $a$ denotes a symmetric $(n+1)\times(n+1)$-matrix whose elements $a^2_{kl}=a^2(\tau_k,\tau_l)$ are obtained from the harmonic Green function for
periodic paths $
G^{\Omega,{\rm p}}(\tau,\tau')$ as (see Chapters 3 and 5 in \cite{PI})
\begin{equation}
  \label{greenh}
  a^2(\tau,\tau')\equiv \frac{\hbar}{M}G^{\Omega,{\rm p}}(\tau,\tau')=\frac{\hbar}{2M\Omega}\frac{\cosh{\Omega(|\tau-\tau'|-\hbar\beta/2)}}{\sinh{\hbar\beta\Omega/2}}.
\end{equation}
The diagonal elements $a^2=a(\tau,\tau)$ represent the fluctuation width (\ref{zmwidQS}) which behaves in the classical limit like (\ref{zmwidcl}) and at zero temperature like (\ref{qmflwidth}).

Both smearing formulas (\ref{smear}) and (\ref{smeard}) allow in principle to determine all harmonic expectation values for the variational perturbation theory of density matrices in terms of ordinary Gaussian integrals. Unfortunately, in many applications containing nonpolynomial potentials, it is impossible to solve neither the spatial nor the temporal integrals analytically. This circumstance drastically increases the numerical effort in higher-order calculations.
\section{First-Order Variational Results}
\label{fstlim}
The first-order variational approximation gives usually a reasonable estimate for any desired quantity. Let us investigate the classical and the quantum mechanical limit of this approximation. To facilitate the discussion, we first derive a new representation for the first-order smearing formula (\ref{smeard}) which allows a direct evaluation of the imaginary time integral. The resulting expression will depend only on temperature, whose low- and high-temperature limits can easily be extracted.
\subsection{Alternative Formula for First-Order Smearing}
\label{newrep}
For simplicity, we restrict ourselves to the case of particle densities and allow only symmetric potentials $V(x)$ centered at the origin. If $V(x)$ has only one minimum at the origin, then also $x_m$ will be zero. If $V(x)$ has several symmetric minima, then $x_m$ goes to zero only at sufficiently high temperatures (see Ref.~\cite{PI}).

To first order, the smearing formula (\ref{smeard}) reads
\begin{equation}
  \label{smearfirst}
  \meand{{\cal A}_{\rm int}[x]}=\frac{1}{\rho_0^\Omega(x_a)}\int\limits_0^{\hbar\beta}d\tau\,\int\limits_{-\infty}^{+\infty}\frac{dz}{2\pi}\,V_{\rm int}(z)\,\frac{1}{\sqrt{a_{00}^2-a_{01}^2}}\exp\left\{-\frac{1}{2}\frac{(z^2+x_a^2)a_{00}-2z x_a a_{01}}{a_{00}^2-a_{01}^2} \right\},
\end{equation}
so that Mehler's summation formula
\begin{equation}
  \label{mehler}
  \frac{1}{\sqrt{1-b^2}}\exp\left\{-\frac{(x^2+x'^2)(1+b^2)-4xx'b}{2(1-b^2)}\right\}=\exp\left\{-\frac{1}{2}(x^2+x'^2) \right\}\sum\limits_{n=0}^\infty\,\frac{b^n}{2^n n!}\,H_n(x) H_n(x')
\end{equation}
leads to an expansion in terms of Hermite polynomials $H_n(x)$, whose temperature dependence stems from the diagonal elements of the harmonic Green function (\ref{greenh}):
\begin{equation}
  \label{fstexp}
  \meand{{\cal A}_{\rm int}[x]}=\sum\limits_{n=0}^\infty\,\frac{\hbar\beta}{2^n n!}\,C^{(n)}_\beta\,H_n\left(x_a/\sqrt{2a^2_{00}}\right)\,\int\limits_{-\infty}^{+\infty}\frac{dz}{\sqrt{2\pi a^2_{00}}}\,V_{\rm int}(z)\,e^{-z^2/2 a^2_{00}}\,H_n\left(z/\sqrt{2 a^2_{00}}\right).
\end{equation}
Here the dimensionless functions $C_\beta^{(n)}$ are defined by
\begin{equation}
  \label{timeint}
  C_\beta^{(n)}=\frac{1}{\hbar\beta}\int\limits_0^{\hbar\beta}d\tau\,\left(\frac{a^2_{01}}{a^2_{00}}\right)^n.
\end{equation}
Inserting (\ref{greenh}) and performing the integral over $\tau$, we obtain
\begin{equation}
  \label{cpoly}
  C_\beta^{(n)}=\frac{1}{2^n\cosh^n\hbar\beta\Omega/2}
\sum\limits_{k=0}^n\,\left(
n\atop k
\right)\,
\frac{\sinh\hbar\beta\Omega(n/2-k)}{\hbar\beta\Omega(n/2-k)}.
\end{equation}
At high temperatures, these functions of $\beta$ go all to unity,
\begin{equation}
\label{polyhigh}
\lim_{\beta\to 0}C_\beta^{(n)}=1,
\end{equation}
whereas at zero temperature:
\begin{equation}
\label{polylow}
\lim_{\beta\to\infty}
C_\beta^{(n)}=\left\{
\begin{array}{cl}
1, &~~~~~~ n=0,
\\
\displaystyle
\frac{2}{\hbar\beta\Omega n},&~~~~~~ n>0.  \\
\end{array}
\right.
\end{equation}
\begin{figure}[t]
\centerline{
\setlength{\unitlength}{1cm}
\begin{picture}(12.0,9.0)
\put(0,0){\makebox(12,9){\epsfxsize=12cm \epsfbox{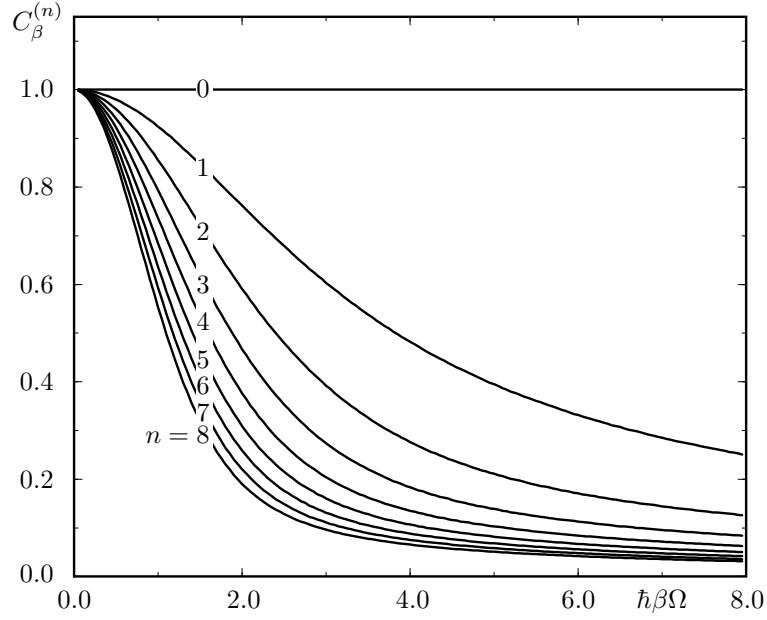}}}
\put(1.1,8.40){$C_\beta^{(n)}$}
\put(9.4,0.7){$\hbar\beta\Omega$}
\put(2.87,2.9){$n=8$}
\put(3.55,3.2){7}
\put(3.55,3.55){6}
\put(3.55,3.9){5}
\put(3.55,4.4){4}
\put(3.55,4.9){3}
\put(3.55,5.6){2}
\put(3.55,6.45){1}
\put(3.55,7.5){0}
\put(1.7,0.7){0.0}
\put(3.95,0.7){2.0}
\put(6.20,0.7){4.0}
\put(8.40,0.7){6.0}
\put(10.65,0.7){8.0}
\put(1.2,1.05){0.0}
\put(1.2,2.3){0.2}
\put(1.2,3.6){0.4}
\put(1.2,4.90){0.6}
\put(1.2,6.20){0.8}
\put(1.2,7.50){1.0}
\end{picture}
}
\caption{\label{cnbeta} Temperature-dependence of the first 9 functions $C_\beta^{(n)}$, where $\beta=1/k_BT$.}
\end{figure}

According to (\ref{veff}), the
first-order approximation to the
new effective potential (\ref{veff}) is given by
\begin{equation}
  \label{Wfirst}
  W_1^\Omega(x_a)=\frac{1}{2\beta}\ln\frac{\sinh{\hbar\beta\Omega}}{\hbar\beta\Omega}+\frac{M\Omega}{\hbar\beta}x_a^2\tanh\frac{\hbar\beta\Omega}{2}+V_{a^2}^\Omega(x_a)
\end{equation}
with the smeared interaction potential
\begin{equation}
  \label{smepot}
  V_{a^2}^\Omega(x_a)=\frac{1}{\hbar\beta}\,\meand{{\cal A}_{\rm int}[x]}.
\end{equation}
It is instructive to discuss separately the limits $\beta\to 0$ and $\beta\to \infty$ of dominating thermal and quantum fluctuations, respectively.
\subsection{Classical Limit of Effective Classical Potential}
\label{clli}
In the classical limit $\beta\to 0$, the first-order
effective classical potential (\ref{Wfirst}) reduces to
\begin{equation}
  \label{Wbzero}
  W_1^{\Omega,{\rm cl}}(x_a)=\frac{1}{2}M\Omega^2x_a^2+\lim_{\beta\to 0}\,V_{a^2}^\Omega(x_a).
\end{equation}
The second term is determined by inserting the high-temperature limit of the fluctuation width (\ref{zmwidcl}) and of the polynomials (\ref{polyhigh}) into the expansion (\ref{fstexp}), leading to
\begin{equation}
  \label{vazeroA}
  \lim_{\beta\to 0}\,V_{a^2}^\Omega(x_a)=\lim_{\beta\to 0}\sum\limits_{n=0}^\infty\,\frac{1}{2^n n!}\,H_n\left(\sqrt{M\Omega^2\beta/2}\,x_a\right)\int\limits_{-\infty}^{+\infty}\frac{dz}{\sqrt{2\pi/M\Omega^2\beta}}\,V_{\rm int}(z)\,e^{-M\Omega^2\beta\,z^2/2}H_n\left(\sqrt{M\Omega^2\beta/2}\,z\right).
\end{equation}
Then we make use of the completeness relation for Hermite polynomials
\begin{equation}
  \label{complherm}
  \frac{1}{\sqrt{\pi}} e^{-x^2}\sum\limits_{n=0}^\infty\,\frac{1}{2^n n!}\,H_n(x)\,H_n(x')=\delta(x-x'),
\end{equation}
which may be derived from Mehler's summation formula (\ref{mehler}) in the limit $b\to 1^-$, to reduce the smeared interaction potential $V_{a^2}^\Omega(x_a)$ to the pure interaction potential (\ref{potint}):
\begin{equation}
  \label{vazeroB}
  \lim_{\beta\to 0}\,V_{a^2}^\Omega(x_a)=V_{\rm int}(x_a).
\end{equation}
Recalling (\ref{potint}) we see that the first-order
effective classical potential (\ref{Wbzero}) approaches the classical one:
\begin{equation}
  \label{effpotzero}
  \lim_{\beta\to 0}\,W_1^{\Omega,{\rm cl}}(x_a)=V(x_a).
\end{equation}
This is a consequence of the vanishing fluctuation width $b_{\rm H}^2$ of the paths around the classical orbits. This property is universal to all higher-order approximations to the
effective classical potential (\ref{veff}). Thus all corrections terms with $n>1$ must disappear in the limit $\beta\to 0$,
\begin{equation}
  \label{highzero}
  \lim_{\beta\to 0}\,\frac{-1}{\beta}\sum_{n=2}^\infty\,\frac{(-1)^n}{n! \hbar^n}\,\cumd{{\cal A}_{\rm int}^n[x]}=0.
\end{equation}
\subsection{Zero-Temperature Limit}
\label{quli}
At low temperatures, the first-order
effective classical potential (\ref{Wfirst}) becomes
\begin{equation}
  \label{quli1}
  W^{\Omega,{\rm qm}}_1(x_a)=\frac{\hbar\Omega}{2}+\lim_{\beta\to\infty}V_{a^2}^\Omega(x_a).
\end{equation}
The zero-temperature limit of the smeared potential
in the second term defined in (\ref{smepot})
follows from Eq.~(\ref{fstexp}) by taking into account the limiting procedure for the polynomials $C_\beta^{(n)}$ in (\ref{polylow}) and for the fluctuation width $a^2_{\rm qm}$ (\ref{qmflwidth}). Thus we obtain with $H_0(x)=1$ and the inverse length $\kappa=\sqrt{M\Omega/\hbar}$:
\begin{equation}
  \label{quli2}
  \lim_{\beta\to\infty} V_{a^2}^\Omega(x_a)=\int\limits_{-\infty}^{+\infty}dz\,\sqrt{\frac{\kappa^2}{\pi}}H_0(\kappa z)^2\exp\{-\kappa^2z^2\}\,V_{\rm int}(z).
\end{equation}
Introducing the harmonic eigenvalues
\begin{equation}
  \label{harmen}
  E^\Omega_n=\hbar\Omega\left(n+\frac{1}{2}\right),
\end{equation}
and the harmonic eigenfunctions
\begin{equation}
  \label{harmwave}
  \psi_n^\Omega(x)=\frac{1}{\sqrt{n! 2^n}}\,\left(\frac{\kappa^2}{\pi}\right)^{1/4}\,e^{-\frac{1}{2}\kappa^2x^2}\,H_n(\kappa x),
\end{equation}
we can reexpress the zero-temperature limit of the first-order
effective classical potential (\ref{quli1}) with (\ref{quli2}) by
\begin{equation}
  \label{quli3}
  W_1^{\Omega,{\rm qm}}(x_a)=E_0^\Omega+\langle\,\psi_0^\Omega\,|\,V_{\rm int}\,|\,\psi_0^\Omega\,\rangle.
\end{equation}
This is recognized as the first-order harmonic Rayleigh-Schr\"odinger perturbative result for the ground state energy.

For the discussion of the quantum mechanical limit of the first-order normalized density,
\begin{equation}
  \label{denszeroA}
  \rho_1^\Omega(x_a)=\frac{\trho_1^{~\Omega}(x_a)}{Z}=\rho_0^\Omega(x_a)\,\frac{\exp\left\{-\frac{1}{\hbar} \meand{{\cal A}_{\rm int}[x]}\right\}}{\int_{-\infty}^{+\infty} dx_a\,
\rho_0^\Omega(x_a)\exp\left\{-\frac{1}{\hbar}\meand{{\cal A}_{\rm int}[x]}\right\}},
\end{equation}
we proceed as follows.
First we expand (\ref{denszeroA}) up to first order in the interaction, leading to
\begin{equation}
  \label{dmfirst1}
  \rho_1^\Omega(x_a)=\rho_0^\Omega(x_a)
\left[1-\frac{1}{\hbar}\left(\meand{{\cal A}_{\rm int}[x]}-\int\limits_{-\infty}^{+\infty}d x_a\,
\rho_0^\Omega(x_a)\,\meand{{\cal A}_{\rm int}[x]} \right) \right].
\end{equation}
Inserting (\ref{displ}) and (\ref{fstexp}) into the third term in (\ref{dmfirst1}), and assuming $\Omega$ not to depend explicitly on $x_a$, the $x_a$-integral reduces to the orthonormality relation for Hermite polynomials
\begin{equation}
  \label{ortho}
  \frac{1}{2^nn!\sqrt{\pi}}\int\limits_{-\infty}^{+\infty}dx_a H_n(x_a)H_0(x_a)e^{-x_a^2}=\delta_{n0},
\end{equation}
so that the third term in (\ref{dmfirst1}) eventually becomes
\begin{equation}
  -\int\limits_{-\infty}^{+\infty}d x_a\,\rho_0^\Omega(x_a)\,\meand{{\cal A}_{\rm int}[x]}=-\beta\int\limits_{-\infty}^{+\infty}dz\,\sqrt{\frac{\kappa^2}{\pi}}\,V_{\rm int}(z)\,\exp\{-\kappa^2z^2\}\,H_0(\kappa z).
\end{equation}
But this is just the $n=0$~-term of (\ref{fstexp}) with an opposite sign, thus cancelling the zeroth component of the second term in (\ref{dmfirst1}), which would have been divergent for $\beta\to\infty$.

The resulting expression for the first-order normalized density is
\begin{equation}
  \label{dmfirst2}
  \rho^\Omega_1(x_a)=\rho_0^\Omega(x_a)\left[1-\sum\limits_{n=1}^\infty\,\frac{\beta}{2^n n!}\,C_\beta^{(n)}\,H_n(\kappa x_a)\int\limits_{-\infty}^{+\infty}dz\,\sqrt{\frac{\kappa^2}{\pi}}\,V_{\rm int}(z)\,\exp(-\kappa^2z^2)\,H_n(\kappa z)\right].
\end{equation}
The zero-temperature limit of $c_\beta^{(n)}$ is from (\ref{polylow}) and (\ref{harmen})
\begin{equation}
  \lim_{\beta\to\infty}\beta C_\beta^{(n)}=\frac{2}{E_n^\Omega-E_0^\Omega},
\end{equation}
so that we obtain from (\ref{dmfirst2}) the limit
\begin{equation}
  \label{dmfirst3}
 \rho_1^\Omega(x_a)=\rho_0^\Omega(x_a)\,\Bigg[1-2\sum\limits_{n=1}^\infty\,\frac{1}{2^n n!}\frac{1}{E_n^\Omega-E_0^\Omega}H_n(\kappa x_a )\int\limits_{-\infty}^{\infty}dz\,\sqrt{\frac{\kappa^2}{\pi}}\,V_{\rm int}(z)\,\exp\{-\kappa^2z^2 \} H_n(\kappa z)\,H_0(\kappa z )\Bigg].
\end{equation}
Taking into account the harmonic eigenfunctions (\ref{harmwave}), we can rewrite (\ref{dmfirst3}) as
\begin{equation}
  \label{dmfirst4}
  \rho_1^\Omega(x_a)=|\psi_0(x_a)|^2=[\psi_0^\Omega(x_a)]^2-2\psi_0^\Omega(x_a)\sum\limits_{n>0}\psi_n^\Omega(x_a)\frac{\langle\,\psi_n^\Omega\,|\,V_{\rm int}\,|\,\psi_0^\Omega\,\rangle}{E_n^\Omega-E_0^\Omega}
\end{equation}
which is just equivalent to the harmonic first-order Rayleigh-Schr\"odinger result for particle densities.

Summarizing the results of this section, we have shown that our method has properly reproduced the high- and low-temperature limits. Because of relation (\ref{dmfirst4}), the variational approach for particle densities can be used to determine approximately the ground state wave function $\psi_0(x_a)$ for the system of interest.
\section{Smearing Formula in Higher Spatial Dimensions}
\label{highdim}
Most physical systems possess many degrees of freedom. This requires an extension of our method to higher spatial dimensions. In general, we must consider anisotropic harmonic trial systems, in which the previous variational parameter $\Omega^2$ becomes a $D\times D$--matrix $\Omega^2_{\mu\nu}$ with $\mu,\nu=1,2,\ldots,D$.
\subsection{Isotropic Approximation}
\label{highiso}
An isotropic trial ansatz
\begin{equation}
  \label{isoOmega}
  \Omega^2_{\mu\nu}=\Omega^2\delta_{\mu\nu}
\end{equation}
can give rough initial estimates for the properties of the system.
In this case, the $n$-th order smearing formula (\ref{smeard}) generalizes directly to
\begin{equation}
  \label{smeariso}
  \meaniso{{\cal A}_{\rm int}^n[{\bf r}]}=\frac{1}{\rho_0^\Omega({\bf r}_a)}\prod\limits_{l=1}^n\left[\int_0^{\hbar\beta}d\tau_l\int d^Dz_l\,V_{\rm int}({\bf z}_l)\right]\,\frac{1}{\sqrt{(2\pi)^{n+1}\,{\rm det}\,a^2}^D}\,\exp\left[-\frac{1}{2}\sum\limits_{k,l=0}^n\,{\bf z}_k\,a^{-2}_{kl}\,{\bf z}_l \right]
\end{equation}
with the $D$--dimensional vectors ${\bf z}_l=(z_{1l},z_{2l},\ldots,z_{Dl})^T$. Note, that greek labels $\mu,\nu,\ldots=1,2,\ldots,D$ specify spatial indices and latin labels $k,l,\ldots=0,1,2,\ldots,n$ refer to the different imaginary times. The vector ${\bf z}_0$ denotes ${\bf r}_a$, the matrix $a^2$ is the same as in Section~\ref{smform}. The harmonic density reads
\begin{equation}
  \label{isoharmdens}
  \rho_0^{\Omega}({\bf r})=\sqrt{\frac{1}{2\pi a^2_{00}}}^D\,\exp\left[-\frac{1}{2\,a^2_{00}}\sum\limits_{\mu=1}^{D}\,x_\mu^2\right].
\end{equation}
\subsection{Anisotropic Approximation}
In the discussion of the anisotropic approximation, we shall consider only radially-symmetric potentials $V({\bf r})=V(|{\bf r}|)$ for simplicity and their major occurence in physics. The trial frequencies decompose naturally into a radial frequency $\Omega_L$ and a transverse one $\Omega_T$ (see Ref.~\cite{PI}):
\begin{equation}
  \label{anisoOmega}
  \Omega^2_{\mu\nu}=\Omega_L^2\,\frac{{x_a}_\mu {x_a}_\nu}{r_a^2}+\Omega^2_T\left(\delta_{\mu\nu}-\frac{{x_a}_\mu {x_a}_\nu}{r_a^2}\right)
\end{equation}
with $r_a=|{\bf r}_a|$. For practical reasons we rotate the coordinate system by $\bar{{\bf x}}_n=U\,{\bf x}_n$ so that $\overline{\bf r}_a$ points along the first coordinate axis,
\begin{equation}
  \label{ra}
  (\bar{\bf r}_a)_\mu\equiv \bar{z}_{\mu 0}=\left\{\begin{array}{cc}r_a, & \mu=1,\\ 0, & 2\le \mu\le D,  \end{array} \right.
\end{equation}
and $\Omega^2$-matrix is diagonal:
\begin{equation}
  \label{rotOmega}
  \overline{\Omega^2}=\left(\begin{array}{ccccc}
\Omega^2_L & 0 & 0 & \cdots & 0\\
0 & \Omega_T^2 & 0 & \cdots & 0\\
0 & 0 &\Omega^2_T & \cdots &0\\
\vdots & \vdots &\vdots &\ddots &\vdots\\
0&0&0&\cdots& \Omega^2_T
\end{array}\right)=U\,\Omega^2\,U^{-1}.
\end{equation}
After this rotation, the {\it anisotropic} $n$-th order smearing formula in $D$ dimensions reads
\begin{eqnarray}
  \label{smearaniso}
\lefteqn{\meananiso{{\cal A}_{\rm int}^n[{\bf r}]}=\frac{1}{
\rho_0^{\Omega_{L,T}}(\bar{\bf r}_a)} \prod\limits_{l=1}^n\left[\int_0^{\hbar\beta}d\tau_l\,\int d^D\bar{z}_l\,V_{\rm int}(|\bar{\bf z}_l|)\right]\,(2\pi)^{-D(n+1)/2}({\rm det}\,a^2_L)^{-1/2}\,({\rm det}\,a^2_T)^{-(D-1)/2}}\hspace{140pt}\nonumber\\
&\times& \exp\left\{-\frac{1}{2}\sum\limits_{k,l=0}^n\,\bar{z}_{1k}{a_L}_{kl}^{-2}\bar{z}_{1l}\right\}\,\exp\left\{-\frac{1}{2}\sum\limits_{\mu=2}^D\sum_{k,l=1}^n\,\bar{z}_{\mu k}{a_T}_{kl}^{-2}\bar{z}_{\mu l} \right\}.
\end{eqnarray}
The components of the longitudinal and transversal matrices $a^2_L$ and $a^2_T$ are
\begin{equation}
  \label{LTmatrix}
  {a^2_{L}}_{kl}=a^2_{L}(\tau_k,\tau_l),\quad {a^2_{T}}_{kl}=a^2_{T}(\tau_k,\tau_l)
\end{equation}
where the frequency $\Omega$ in (\ref{greenh}) must be substituted by the new variational parameters $\Omega_L,\Omega_T$, respectively. For the harmonic density in the rotated system we find
\begin{equation}
  \label{LTharmdens}
  \rho_0^{\Omega_{L,T}}(\bar{\bf r})=\sqrt{\frac{1}{2\pi {a^2_L}_{00}}}\,\sqrt{\frac{1}{2\pi {a^2_T}_{00}}}^{D-1}\,\exp\left[-\frac{1}{2\, {a^2_L}_{00}}\bar{x}_1^2-\frac{1}{2\,{a^2_T}_{00}}\sum\limits_{\mu=2}^{D}\bar{x}_\mu^2\right]
\end{equation}
which is used to normalize (\ref{smearaniso}).

The anisotropic smearing formula (\ref{smearaniso}) will be applied to the Coulomb problem below. The anisotropy becomes significant only at low temperatures, where radial and transversal quantum fluctuations have quite different weights. The
effect of anisotropy disappears completely in the classical limit.
\section{Applications}
\label{applics}
In this section we apply the theory to calculate the electron density of the hydrogen atom.
For simplicity, we
shall employ
	 natural units with
$\hbar=k_B=M=1$. In order to
develop some feeling how the approximations work,
we first determine the particle density in
a double-well potential to second order.
\subsection{The Double-Well}
\label{dwsect}
\begin{figure}
\cm{\noindent\parbox{10cm}{
\parbox{3cm}{{\bf a)} $x_a=-\displaystyle{\frac{1}{\sqrt{g}}}$}\hfill
\parbox{7cm}{
\centerline{
\setlength{\unitlength}{1cm}
\begin{picture}(8,6.5)
\put(0,0){\makebox(8,6.5){\epsfxsize=6.5 cm \epsfbox{./w3_3d.eps}}}
\put(0.3,2.6){$1.0$}
\put(0.5,2.3){$0.8$}
\put(0.7,2.0){$0.6$}
\put(0.2,3.1){$W_1$}
\put(3.4,0.8){$0.0$}
\put(1.9,1.3){$-1.0$}
\put(4.6,0.3){$x_m$}
\put(6.9,4.3){$0.5$}
\put(6.5,2.9){$1.5$}
\put(6.0,1.0){$2.5$}
\put(6.3,2.0){$\Omega^2$}
\end{picture}
}}}\hfill
\parbox{7cm}{
\centerline{
\setlength{\unitlength}{1cm}
\begin{picture}(8,6.5)
\put(0,0){\makebox(8,6.5){\epsfxsize=6 cm \epsfbox{./w3_dens.eps}}}
\put(1.2,0.1){$0.0$}
\put(3.95,0.1){$3.0$}
\put(6.7,0.1){$6.0$}
\put(5.45,0.1){$\Omega^2$}
\put(0.7,3.3){$0.0$}
\put(0.4,0.5){$-2.0$}
\put(0.7,6.0){$2.0$}
\put(0.7,4.75){$x_m$}
\end{picture}
}}\\ }
\parbox{10cm}{
\parbox{3cm}{}\hfill
\parbox{7cm}{
\centerline{
\setlength{\unitlength}{1cm}
\begin{picture}(8,6.5)
\put(0,0){\makebox(8,6.5){\epsfxsize=6.5 cm \epsfbox{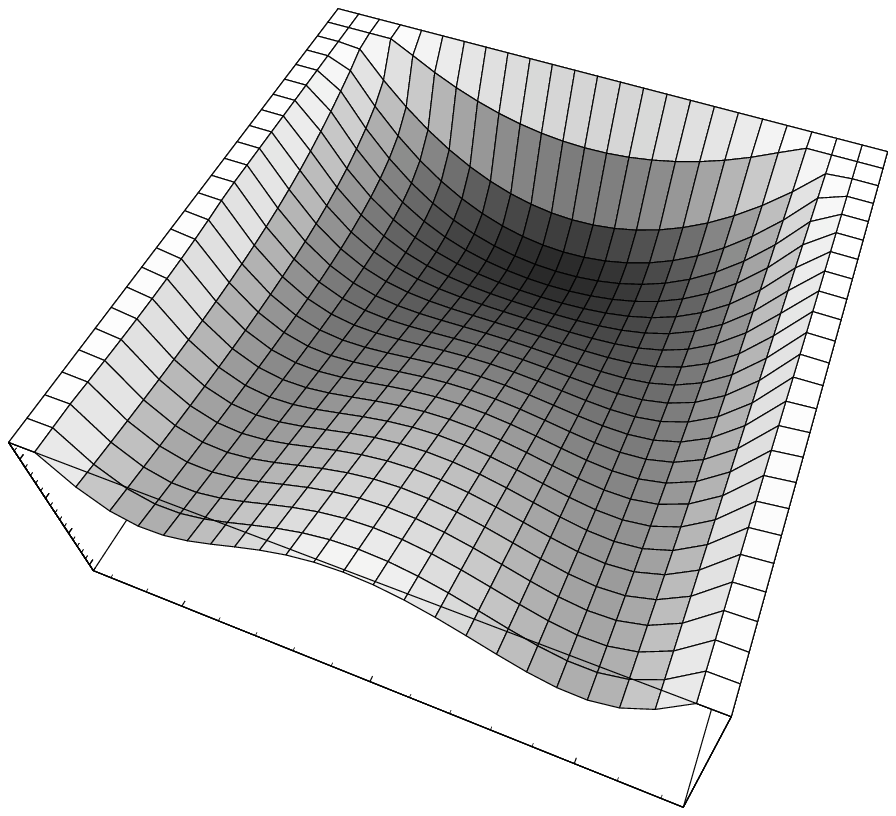}}}
\put(0.4,5.5){{\bf a)} $x_a=0$}
\put(0.5,2.8){$0.6$}
\put(0.9,2.2){$0.5$}
\put(0.4,3.3){$W_1$}
\put(3.1,1.1){$0.0$}
\put(1.7,1.5){$-1.0$}
\put(4.4,0.6){$x_m$}
\put(6.2,1.9){$1.0$}
\put(6.5,2.8){$\Omega^2$}
\put(6.7,3.55){$0.5$}
\put(6.8,4.2){$0.25$}
\end{picture}
}}}\hfill
\cm{\parbox{7cm}{
\centerline{
\setlength{\unitlength}{1cm}
\begin{picture}(8,6.5)
\put(-0.07,0){\makebox(8,6.5){\epsfxsize=6.36 cm \epsfbox{./w1_dens.eps}}}
\put(1.23,0.1){$0.0$}
\put(3.98,0.1){$0.6$}
\put(6.73,0.1){$1.2$}
\put(5.43,0.1){$\Omega^2$}
\put(0.73,3.3){$0.0$}
\put(0.43,0.5){$-1.5$}
\put(0.73,6.1){$1.5$}
\put(0.73,4.75){$x_m$}
\end{picture}
}}}
\phantom{xx}\parbox{10cm}{
\parbox{3cm}{}\hfill
\parbox{7cm}{
\centerline{
\setlength{\unitlength}{1cm}
\begin{picture}(8,6.5)
\put(0,0){\makebox(8,6.5){\epsfxsize=6.5 cm \epsfbox{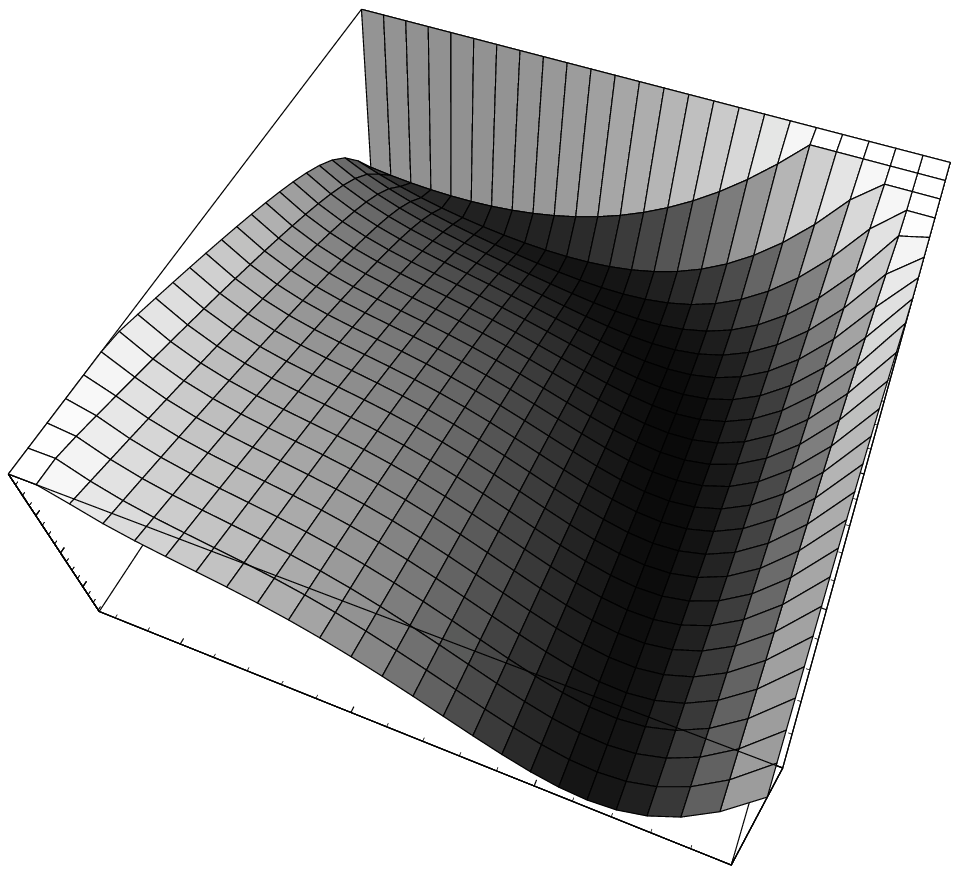}}}
\put(0.2,5.5){{\bf b)} $x_a=\displaystyle{\frac{1}{\sqrt{g}}}$}
\put(0.3,2.6){$1.0$}
\put(0.5,2.3){$0.8$}
\put(0.7,2.0){$0.6$}
\put(0.2,3.1){$W_1$}
\put(2.8,1.2){$0.0$}
\put(1.4,1.6){$-1.0$}
\put(4.0,0.8){$x_m$}
\put(5.2,0.3){$2.0$}
\put(6.9,4.4){$0.5$}
\put(6.5,3.0){$1.5$}
\put(6.0,1.1){$2.5$}
\put(6.3,2.1){$\Omega^2$}
\end{picture}
}}}\hfill
\cm{\parbox{7cm}{
\centerline{
\setlength{\unitlength}{1cm}
\begin{picture}(8,6.5)
\put(0,0){\makebox(8,6.5){\epsfxsize=6 cm \epsfbox{./w2_dens.eps}}}
\put(1.2,0.1){$0.0$}
\put(3.95,0.1){$3.0$}
\put(6.7,0.1){$6.0$}
\put(5.45,0.1){$\Omega^2$}
\put(0.7,3.3){$0.0$}
\put(0.4,0.5){$-2.0$}
\put(0.7,6.0){$2.0$}
\put(0.7,4.75){$x_m$}}
\end{picture}
}}
\caption{\label{w3dplot}  Plots
of the first-order approximation
$W_1^{\Omega,x_m}(x_a)$
to
effective classical potential
as a function of the two
variational parameters $\Omega^2(x_a), x_m(x_a)$
at $g=0.4$ and $\beta=10$ for two different values of $x_a$.
}
\end{figure}

\begin{figure}[t]
\centerline{
\setlength{\unitlength}{1cm}
\begin{picture}(12.0,9.0)
\put(0,0){\makebox(12,9){\epsfxsize=12cm \epsfbox{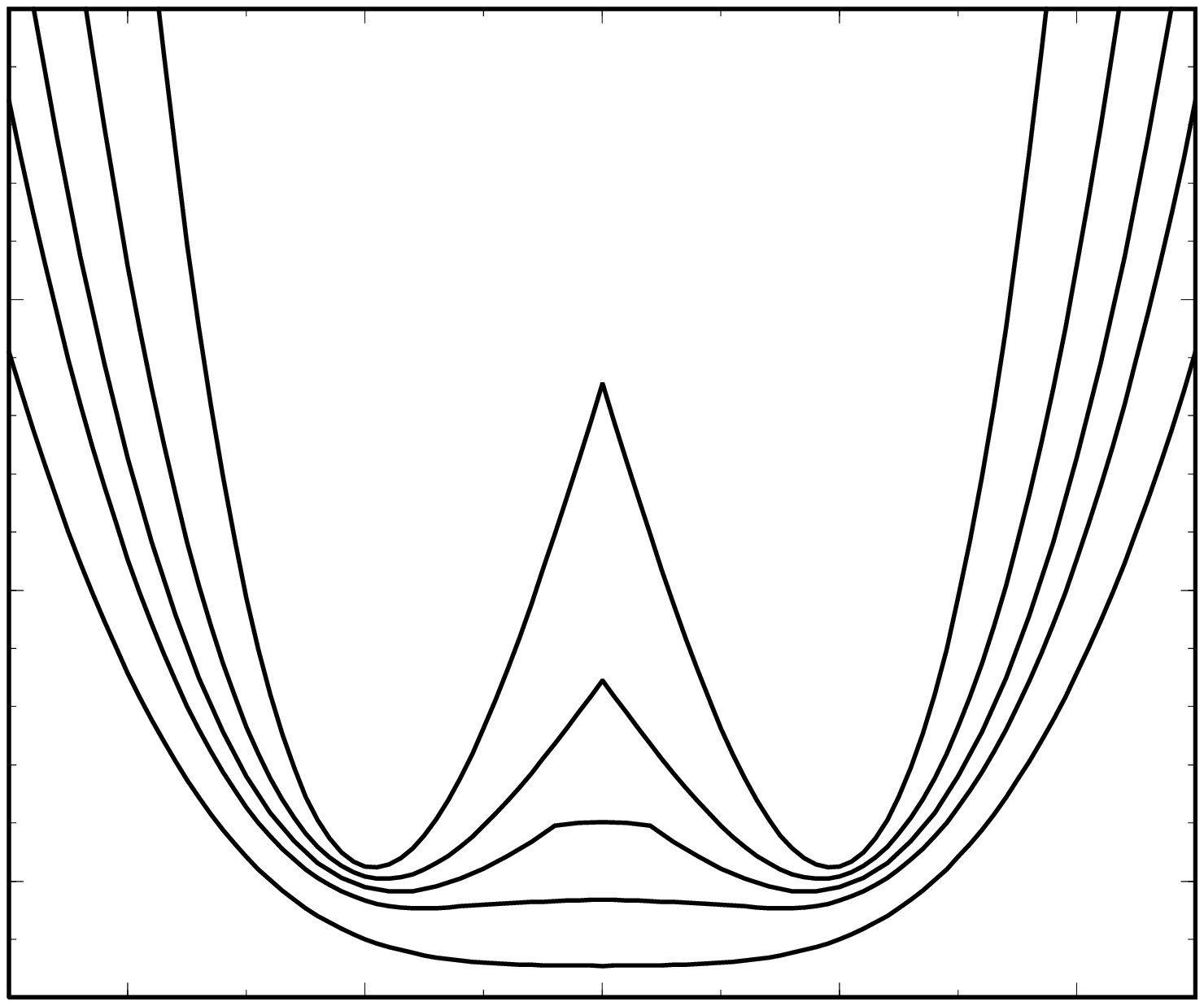}}}
\put(2.2,0.7){$-2/\sqrt{g}$}
\put(4.0,0.7){$-1/\sqrt{g}$}
\put(6.3,0.7){$0$}
\put(7.6,0.7){$+1/\sqrt{g}$}
\put(9.4,0.7){$+2/\sqrt{g}$}
\put(6.2,0.3){$x_a$}
\put(1.3,1.9){$0.5$}
\put(1.3,4.1){$1.0$}
\put(1.3,6.3){$1.5$}
\put(0.7,8.5){$W_1(x_a)$}
\put(5.0,5.4){$g=0.1$}
\put(6.15,3.7){$0.2$}
\put(6.15,2.6){$0.3$}
\put(6.15,2.0){$0.4$}
\put(6.15,1.5){$0.6$}
\end{picture}
}
\caption{\label{w1} First-order approximation of effective classical potential $W_1(x_a)$ for different couplings $g$ as a function of the position $x_a$ at $\beta=10$.}
\end{figure}

\begin{figure}
\centerline{
\setlength{\unitlength}{1cm}
\begin{picture}(12.0,9.0)
\put(0,0){\makebox(12,9){\epsfxsize=12cm \epsfbox{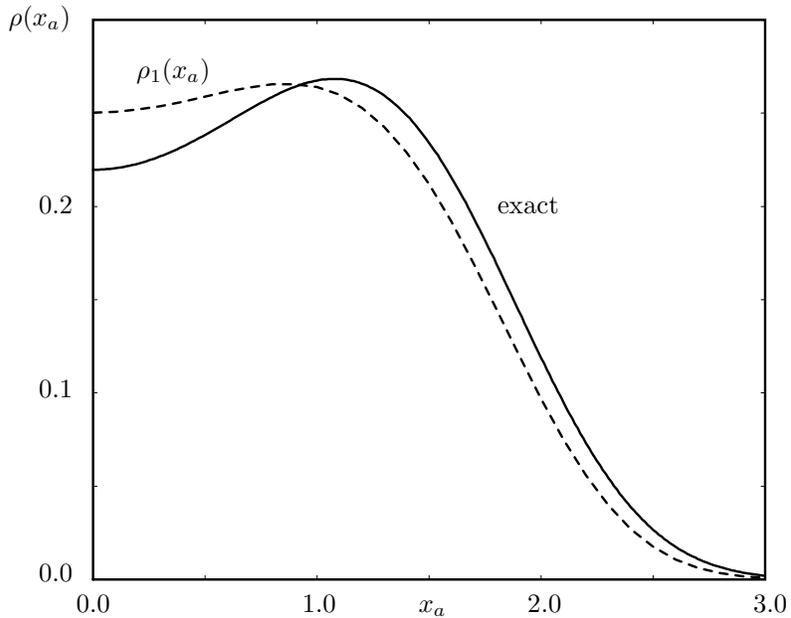}}}
\put(6.25,0.7){$x_a$}
\put(0.8,8.5){$\rho(x_a)$}
\put(7.3,6.0){exact}
\put(2.5,7.8){$\rho_1(x_a)$}
\put(1.7,0.7){0.0}
\put(4.7,0.7){1.0}
\put(7.7,0.7){2.0}
\put(10.7,0.7){3.0}
\put(1.2,1.1){0.0}
\put(1.2,3.55){0.1}
\put(1.2,6.0){0.2}
\end{picture}
}
\caption{\label{dens10} First-order approximation using different for $\beta=10$ and $g=0.4$ compared
with the exact particle density in a double-well from numerical evaluation of Schr\"odinger equation.
All values are in natural units.}
\end{figure}

\begin{figure}
\centerline{
\setlength{\unitlength}{1cm}
\begin{picture}(12.0,9.0)
\put(0,0){\makebox(12,9){\epsfxsize=12cm \epsfbox{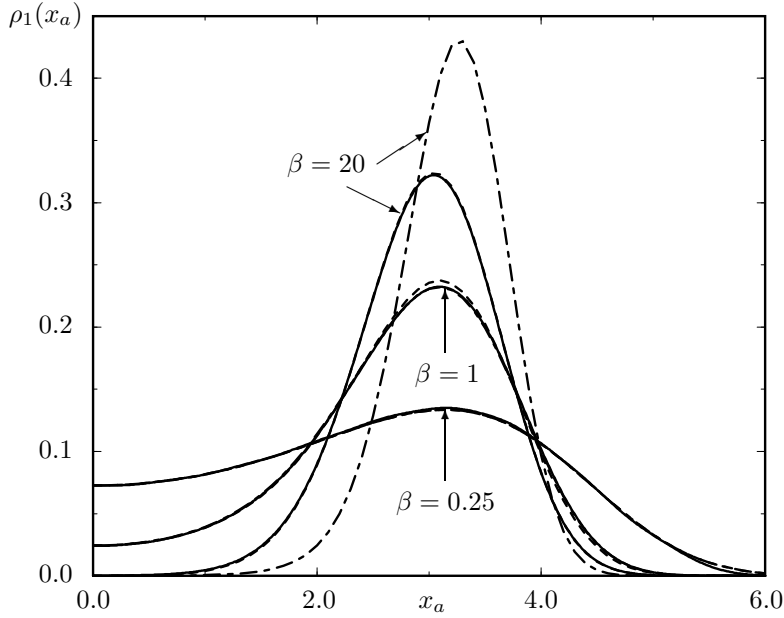}}}
\put(0.8,8.5){$\rho_1(x_a)$}
\put(6.25,0.7){$x_a$}
\put(4.5,6.5){$\beta=20$}
\put(5.3,6.3){\vector(2,-1){0.7}}
\put(5.7,6.6){\vector(3,2){0.65}}
\put(5.95,2){$\beta=0.25$}
\put(6.6,2.4){\vector(0,1){0.95}}
\put(6.2,3.7){$\beta=1$}
\put(6.6,4.1){\vector(0,1){0.85}}
\put(1.7,0.7){0.0}
\put(4.7,0.7){2.0}
\put(7.65,0.7){4.0}
\put(10.65,0.7){6.0}
\put(1.2,1.05){0.0}
\put(1.2,2.7){0.1}
\put(1.2,4.35){0.2}
\put(1.2,6.0){0.3}
\put(1.2,7.65){0.4}
\end{picture}
}
\caption{\label{secvar} First-order particle densities of the double-well for $g=0.1$ obtained by optimizing in two variational parameters $\Omega^2,x_m$ (dashed curves) and with only $\Omega^2$ (dash-dotted) vs. exact distributions (solid) for different temperatures. The parameter $x_m$ is very important for low temperatures.}
\end{figure}

In the case of the double-well potential
\begin{equation}
  \label{dwell}
  V(x)=-\frac{1}{2}\omega^2x^2+\frac{1}{4}g x^4+\frac{1}{4g}
\end{equation}
with coupling constant $g$, we obtain for the expectation of the interaction (\ref{fstexp}) to first order, also setting $\omega^2=1$,
\begin{eqnarray}
  \label{smeartwo}
  \meandA{{\cal A}_{\rm int}[x]}&=&\frac{1}{2}\beta g_0+\frac{1}{2}g_1C_\beta^{(1)}H_1\left((x_a-x_m)/\sqrt{2 a_{00}^2}\right)+\frac{1}{4}g_2 C_\beta^{(2)}H_2\left((x_a-x_m)/\sqrt{2 a_{00}^2}\right)\nonumber\\
& &+\frac{1}{8}g_3C_\beta^{(3)}H_3\left((x_a-x_m)/\sqrt{2 a_{00}^2}\right)+\frac{1}{16}g_4C_\beta^{(4)}H_4\left((x_a-x_m)/\sqrt{2 a_{00}^2}\right)
\end{eqnarray}
with
\begin{eqnarray}
  \label{abbtwo}
  g_0&=&-a_{00}^2(\Omega^2+1)+\frac{3}{2}g a_{00}^4+3g a_{00}^2 x_m^2+\frac{1}{2}gx_m^4+\frac{1}{2g}-\frac{1}{2}x_m^2\nonumber\\
g_1&=&-\sqrt{2a_{00}^2}x_m+\frac{3}{4}g(2 a_{00}^2)^{3/2}x_m+g\sqrt{2 a_{00}^2}x_m^3\nonumber\\
g_2&=&-a_{00}^2(\Omega^2+1)+3ga_{00}^4+3ga_{00}^2x_m^2\nonumber\\
g_3&=&g(2a_{00}^2)^{3/2}x_m\nonumber\\
g_4&=&ga_{00}^4\nonumber.
\end{eqnarray}
Inserting (\ref{smeartwo}) in (\ref{smepot}), we obtain the unnormalized double-well density
\begin{equation}
  \label{densdwAtwo}
  \trho_1^{~\Omega,x_m}(x_a)=\frac{1}{\sqrt{2\pi\beta}}\exp[-\beta W^{\Omega,x_m}_1(x_a)]
\end{equation}
with the first-order
effective classical potential
\begin{equation}
  \label{densW1two}
  W^{\Omega,x_m}_1(x_a)=\frac{1}{2}\ln\frac{\sinh{\beta\Omega}}{\beta\Omega}+\frac{\Omega}{\beta}(x_a-x_m)^2\tanh\frac{\beta\Omega}{2}+\frac{1}{\beta}\,\meandA{{\cal A}_{\rm int}[x]}.
\end{equation}
After optimizing $W_1^{\Omega,x_m}(x_a)$, the normalized first-order
particle density $\rho_1(x_a)$ is found by dividing $\trho_1(x_a)$
by the first-order partition function
\begin{equation}
\label{partfunc1}
 Z_1=\frac{1}{\sqrt{2\pi\beta}}\int\limits_{-\infty}^{+\infty}dx_a\,\exp[-\beta
W_1(x_a)].
\end{equation}
Subjecting $W_1^{~\Omega,x_m}  (x_a)$
to the extremality conditions (\ref{mincond}), we
obtain optimal values for $ \Omega^2(x_a) $ and $x_m(x_a)$.
Usually there is a unique minimum,
but sometimes
this does not exist and a turning point or a
vanishing higher derivative must be used for optimization.
Fortunately, the first case is often realized. Fig.~\ref{w3dplot}
shows the dependence of the first-order
effective classical potential $W^{\Omega,x_m}_1(x_a)$ at $\beta=10$ and $g=0.4$ for three fixed values of position $x_a$ as a function of the variational parameters $\Omega^2(x_a)$ and $x_m(x_a)$ in a three-dimensional plot and its corresponding density plot. Thereby in both representations, the darker the region the smaller the value of $W_1^{\Omega,x_m}$. We can distinguish between deep valleys (darkgray), in which the global minimum resides, and hills (lightgray).
After  having determined roughly
the area around the expected minimum, one
solves numerically the extremality conditions (\ref{mincond})
with some nearby  starting values, to find
the
exact locations of the minimum. The example in Fig.~\ref{w3dplot} gives an
impression of the general features of the
minimization process. First we note that for symmetry
reasons,
\begin{equation}
\label{xmprop}
x_m(x_a)=-x_m(-x_a),
\end{equation}
and
\begin{equation}
  \label{omprop}
  \Omega^2(x_a)=\Omega^2(-x_a).
\end{equation}
Some first-order approximations
to the effective classical potential
$W_1(x_a)$ are shown in Fig.~\ref{w1}
obtained by optimizing in  $\Omega^2(x_a)$ and $x_m(x_a)$.
The sharp maximum ocurring for weak-coupling
is a consequence of the reflection
property (\ref{xmprop})
enforcing a vanishing $x_m(x_a=0)$.
In the strong-coupling regime, on the other hand,
where $x_m(x_a=0)\approx 0$,  the sharp top is absent.
This behaviour is illustrated in the right-hand parts
of Figs.~\ref{omandxm01} and \ref{omandxm10} at different temperatures.
\newcommand{\mblx}{8}
\newcommand{\mbly}{7}
\begin{figure}
\parbox{8.5cm}{
\parbox{0.5cm}{{\bf a)}}\hfill
\parbox{8cm}{
\centerline{
\setlength{\unitlength}{1cm}
\begin{picture}(\mblx,\mbly)
\put(0,0){\makebox(\mblx,\mbly){\epsfxsize=\mblx cm \epsfbox{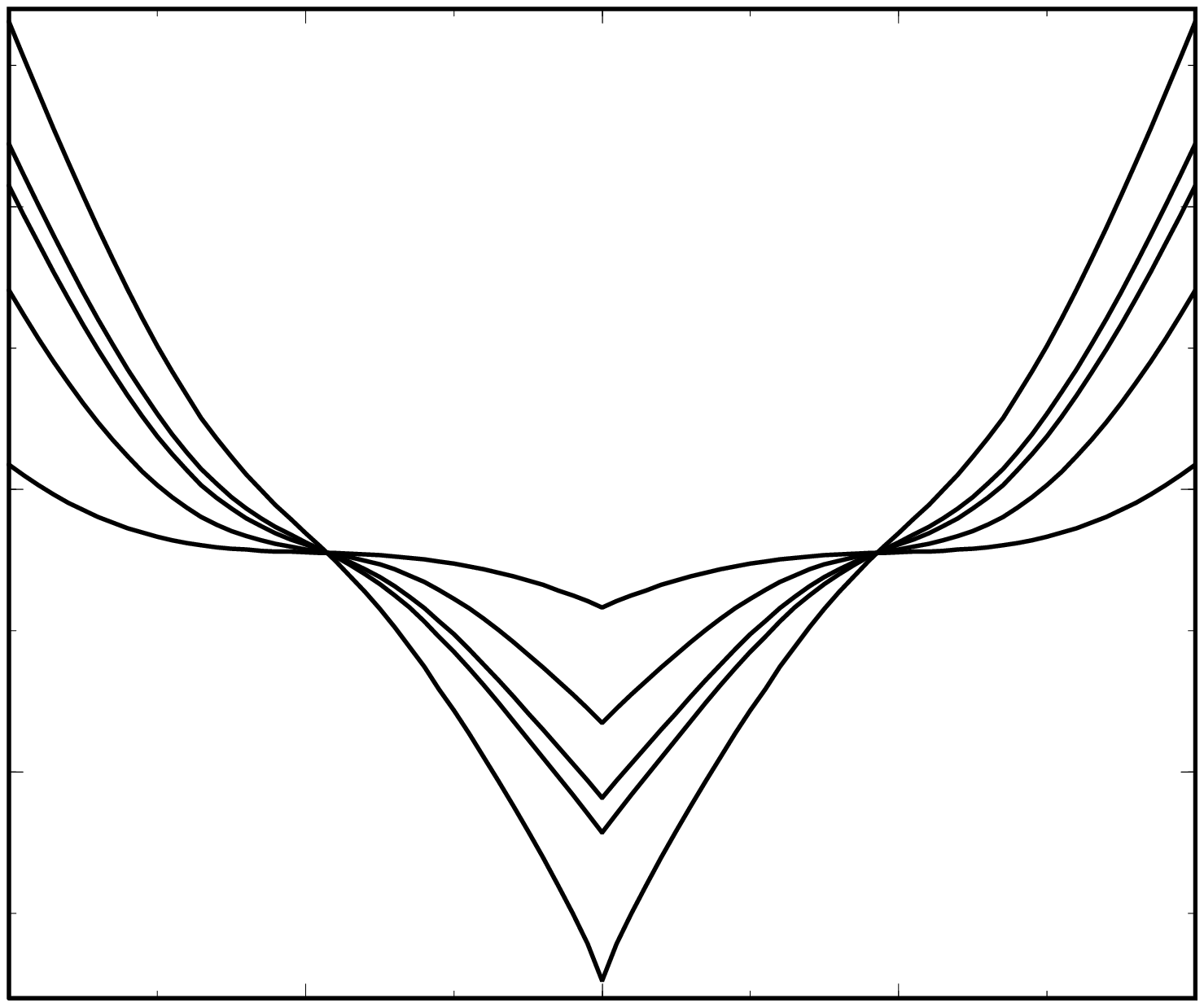}}}
\put(2.1,0.8){$-1/\sqrt{g}$}
\put(4.2,0.8){$0$}
\put(5.15,0.8){$+1/\sqrt{g}$}
\put(4.1,0.5){$x_a$}
\put(0.7,2.3){$1.0$}
\put(0.7,3.7){$2.0$}
\put(0.7,5.1){$3.0$}
\put(0.1,6.1){$\Omega^2(x_a)$}
\put(2.1,5.0){$\beta=5$}
\put(2.8,4.5){$8$}
\put(2.7,4.6){\vector(-1,0){0.9}}
\put(2.6,4){$10$}
\put(2.5,4.1){\vector(-1,0){0.5}}
\put(3.42,3.7){$20$}
\put(3.6,3.6){\vector(0,-1){0.4}}
\put(4.4,3.6){$100$}
\end{picture}
}}}\hfill
\parbox{8.5cm}{
\parbox{0.5cm}{{\bf b)}}\hfill
\parbox{8cm}{
\centerline{
\setlength{\unitlength}{1cm}
\begin{picture}(\mblx,\mbly)
\put(0,0){\makebox(\mblx,\mbly){\epsfxsize=\mblx cm \epsfbox{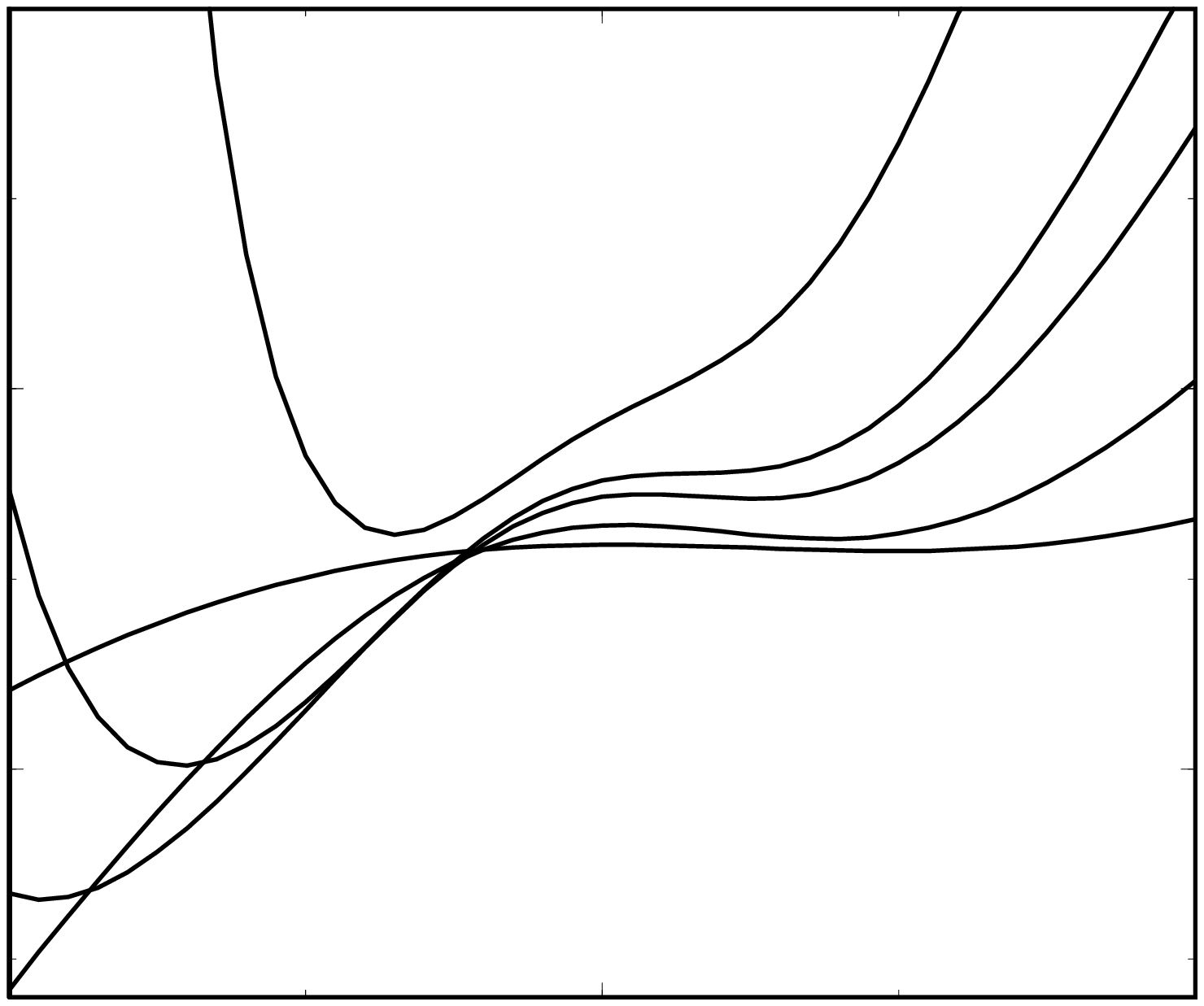}}}
\put(1.2,0.8){$0$}
\put(3.8,0.8){$1/\sqrt{g}$}
\put(6.8,0.8){$2/\sqrt{g}$}
\put(4.1,0.4){$x_a$}
\put(0.5,2.3){$2.95$}
\put(0.5,4.2){$3.00$}
\put(0.1,6.15){$x_m(x_a)$}
\put(2.7,5.0){$\beta=5$}
\put(1.5,3.6){$8$}
\put(2.0,3.4){$100$}
\put(2.3,1.9){$10$}
\put(1.7,1.4){$20$}
\end{picture}
}}}
\caption{\label{omandxm01} {\bf a)} Trial frequency $\Omega^2(x_a)$ at different temperatures and coupling strength $g=0.1$. {\bf b)} Minimum of trial oscillator $x_m(x_a)$
at different temperatures and coupling $g=0.1$.}
\end{figure}
\begin{figure}
\parbox{8.5cm}{
\parbox{0.5cm}{{\bf a)}}\hfill
\parbox{8cm}{
\centerline{
\setlength{\unitlength}{1cm}
\begin{picture}(\mblx,\mbly)
\put(0,0){\makebox(\mblx,\mbly){\epsfxsize=\mblx cm \epsfbox{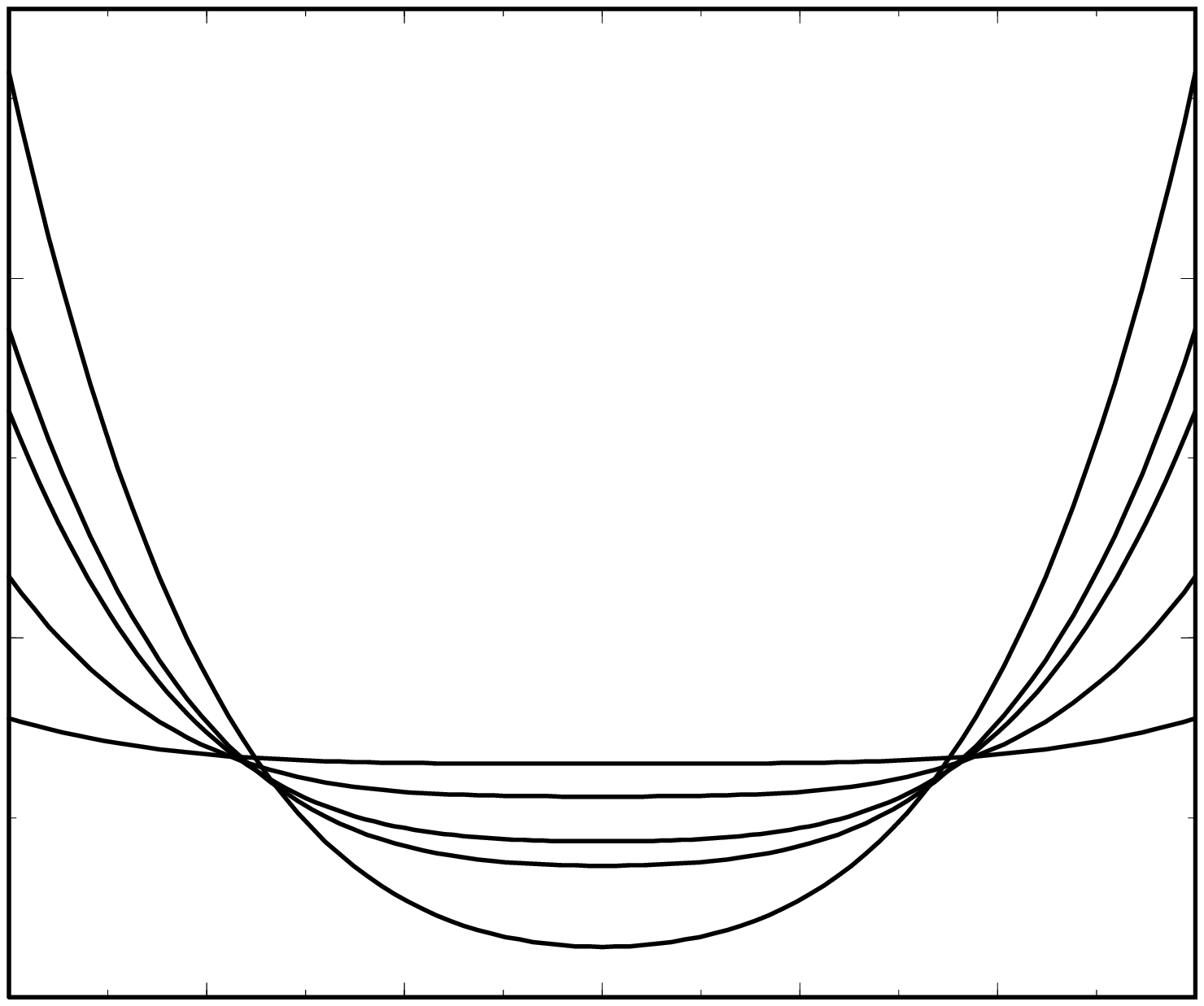}}}
\put(1.6,0.8){$-2/\sqrt{g}$}
\put(4.15,0.8){$0$}
\put(5.8,0.8){$2/\sqrt{g}$}
\put(4.1,0.4){$x_a$}
\put(0.7,1.15){$5.3$}
\put(0.7,2.95){$5.4$}
\put(0.7,4.75){$5.5$}
\put(0.1,6.15){$\Omega^2(x_a)$}
\put(1.7,5.0){$\beta=5$}
\put(2.4,4.0){$8$}
\put(2.3,4.1){\vector(-1,0){0.83}}
\put(2.5,3.0){$10$}
\put(2.4,3.1){\vector(-1,0){0.55}}
\put(1.6,1.9){$20$}
\put(1.75,2.2){\vector(0,1){0.65}}
\put(4.0,2.6){$100$}
\end{picture}
}}}\hfill
\parbox{8.5cm}{
\parbox{0.5cm}{{\bf b)}}\hfill
\parbox{8cm}{
\centerline{
\setlength{\unitlength}{1cm}
\begin{picture}(\mblx,\mbly)
\put(0,0){\makebox(\mblx,\mbly){\epsfxsize=\mblx cm \epsfbox{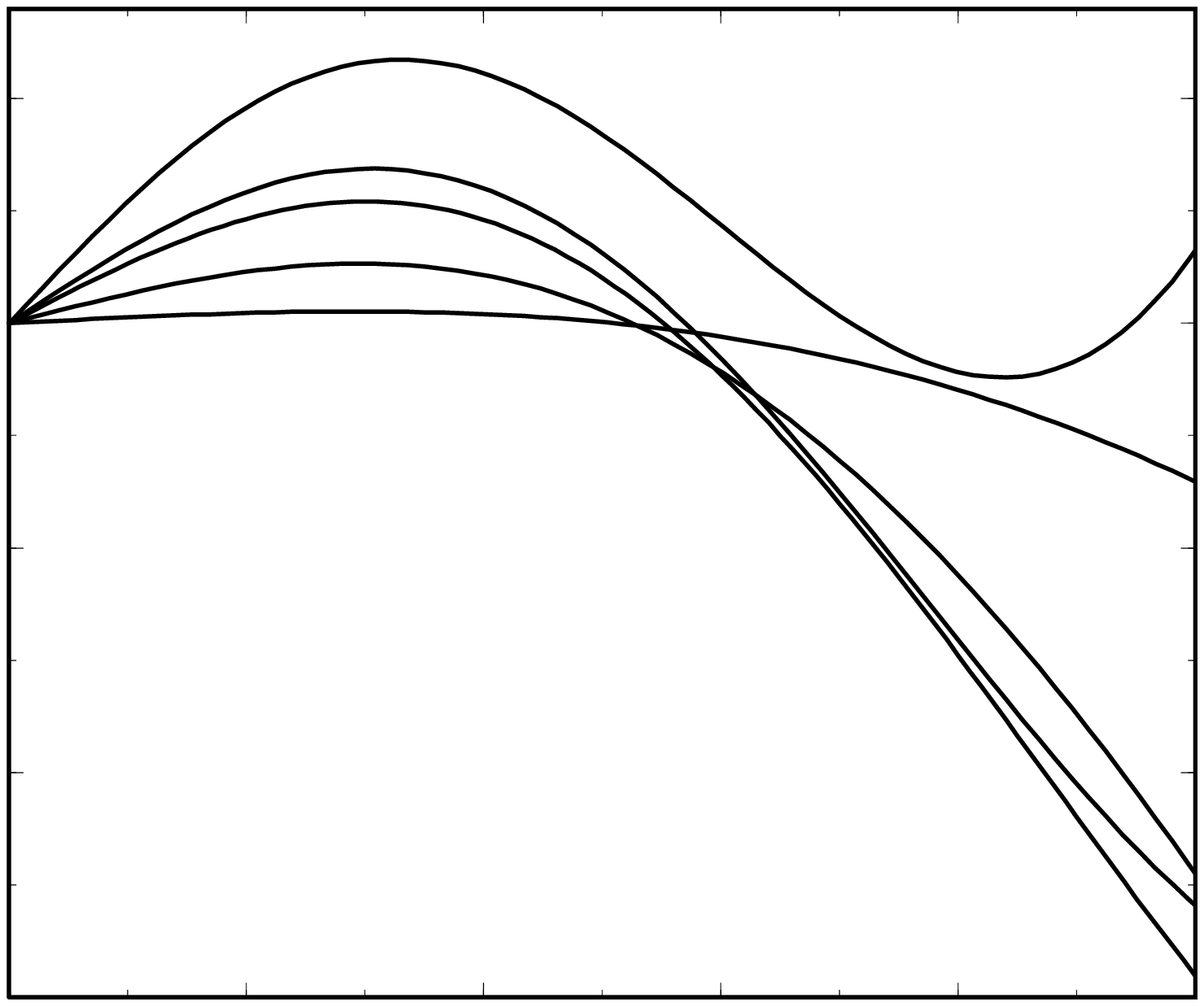}}}
\put(1.2,0.8){$0$}
\put(3.2,0.8){$2/\sqrt{g}$}
\put(5.6,0.8){$4/\sqrt{g}$}
\put(4.1,0.4){$x_a$}
\put(0.25,2.25){$-0.01$}
\put(0.5,4.55){$0.00$}
\put(0.1,6.15){$x_m(x_a)$}
\put(5.0,5.2){$\beta=5$}
\put(3.4,5.5){$8$}
\put(5.1,3.3){$10$}
\put(6.3,3.3){$20$}
\put(2.8,4.3){$100$}
\end{picture}
}}}
\caption{\label{omandxm10} {\bf a)} Trial frequency $\Omega^2(x_a)$ at different temperatures and coupling strength $g=10$. {\bf b)} Minimum of trial oscillator $x_m(x_a)$
at different temperatures and coupling $g=10$.}
\end{figure}
The influence of the center parameter $x_m$ diminuishes
for for increasing values of $g$
and decreasing height $1/4g$ of the central barrier.
The same thing is true
at high temperatures and large values of $x_a$,
where
the precise knowledge of the optimal value of $x_m$ is irrelevant.
In these limits,
the particle density can be determined without optimizing in $x_m$,
setting simply $x_m=0$, where
the expectation value Eq.~(\ref{smeartwo}) reduces to
\begin{equation}
  \label{smeardw}
  \meand{{\cal A}_{\rm int}[x]}=\frac{1}{4}C_\beta^{(2)}H_2\left(x_a/\sqrt{2a_{00}^2}\right)(g_1+3g_2)+\frac{1}{16}\,g_2\, C_\beta^{(4)}H_4\left(x_a/\sqrt{2 a_{00}^2}\right)+\beta\left(\frac{1}{2}g_1+\frac{3}{4}g_2 +g_3\right),
\end{equation}
with
 the abbreviations
\begin{equation}
  \label{abbdw}
  g_1=-a_{00}^2(\Omega^2+1),\quad g_2=g a_{00}^4,\quad g_3=\frac{1}{4g}.
\end{equation}
Inserting (\ref{smeardw}) in (\ref{smepot}) we obtain the unnormalized double-well density
\begin{equation}
  \label{densdwA}
  \trho_1^{~\Omega}(x_a)=\frac{1}{\sqrt{2\pi\beta}}\exp[-\beta W^\Omega_1(x_a)]
\end{equation}
with the first-order effective classical potential
\begin{equation}
  \label{densW1}
  W^\Omega_1(x_a)=\frac{1}{2}\ln\frac{\sinh{\beta\Omega}}{\beta\Omega}+\frac{\Omega}{\beta}x_a^2\tanh\frac{\beta\Omega}{2}+\frac{1}{\beta}\,\meand{{\cal A}_{\rm int}[x]}.
\end{equation}
The
 optimization at $x_m=0$
gives reasonable results
for moderate  temperatures at couplings as low as $g=0.4$,
as shown in Fig.~\ref{dens10} by a
 comparison with the exact density obtained from
numerical solutions of Schr\"odinger equation.
An additional optimization in $x_m$ cannot be distinguished on the plot.
 An example where the second variational parameter $x_m$
does become important is shown in Fig.~\ref{secvar}, where
e we compare the first-order approximation
with one ($\Omega$) and two variational parameters ($\Omega,x_m$)
with the exact density for different temperatures
at the smaller coupling strength $g=0.1$.
 In Fig.~\ref{omandxm01} we see that for $x_a>0$, the optimal $x_m$-values lie close to the right hand minimum of the double-well potential, which we only want to consider here. The minimum is located at $1/\sqrt{g}\approx 3.16$. We observe, that
with two variational parameters the first-order approximation is nearly exact for all temperatures, in contrast to the results with only one variational parameter at low temperatures (see the curve for $\beta=20$).
Also
for a small valuein the case $g=0.4$, the optimization in $\Omega^2$ only gives reasonable results in the temperature region away from high- and low-temperature limits, as shown in Fig.~\ref{dens10} in comparison with the exact density obtained from numerical solution of Schr\"odinger equation. An optimization in two variational parameters gives no better result. A very instructive example, where the second variational parameter $x_m$ becomes important is shown in Fig.~\ref{secvar}. There, we compare the first-order approximation for the double-well density with one ($\Omega$) and two variational parameters ($\Omega,x_m$)
with the exact density for different temperatures at a coupling strength $g=0.1$. This value of $g$ no longer allows to neglect $x_m$ as in the case $g=0.4$. We see from Fig.~\ref{omandxm01} that for $x_a>0$, the optimal $x_m$-values lie close to the right hand minimum of the double-well potential, which we only want to consider here. The minimum is located at $1/\sqrt{g}\approx 3.16$. We observe, that
with two variational parameters the first-order approximation is nearly exact for all temperatures, in contrast to the results with only one variational parameter at low temperatures (see the curve for $\beta=20$).
In second-order variational perturbation theory, the differences between the optimization procedures using one or two variational parameters become less significant. Thus, we restrict ourselves to the optimization in $\Omega(x_a)$.

The second-order density
\begin{equation}
\label{dwdens2}
\trho_2^{~\Omega}(x_a)=\frac{1}{\sqrt{2\pi\beta}}\,\exp[-\beta\,W^\Omega_2(x_a)]
\end{equation}
with the second-order approximation of the effective classical potential
\begin{equation}
W^\Omega_2(x_a)=\frac{1}{2}\ln\frac{\sinh{\beta\Omega}}{\beta\Omega}+\frac{\Omega}{\beta}x_a^2\tanh\frac{\beta\Omega}{2}+\frac{1}{\beta}\,\meand{{\cal A}_{\rm int}[x]}-\frac{1}{2\beta}\,\cumd{{\cal A}_{\rm int}^2[x]}
\end{equation}
requires evaluating the smearing formula (\ref{smear}) for $n=1$ which is given in (\ref{smeardw}) and $n=2$ to be calculated. Going immediately to the cumulant we have
\begin{eqnarray}
  \label{dwcum}
  \cumd{{\cal A}_{\rm int}^2[x]}=\int\limits_0^{\hbar\beta}d\tau_1\int\limits_0^{\hbar\beta}d\tau_2\,&& \Bigg\{ \frac{1}{4}(\Omega^2+1)^2\left[I_{22}(\tau_1,\tau_2)-I_2(\tau_1)I_2(\tau_2) \right]-\frac{1}{4}g(\Omega^2+1)\left[I_{24}(\tau_1,\tau_2)-I_2(\tau_1)I_4(\tau_2) \right] \nonumber\\
& &+\frac{1}{16}g^2\left[I_{44}(\tau_1,\tau_2)-I_4(\tau_1)I_4(\tau_2) \right]\Bigg\}
\end{eqnarray}
with
\begin{equation}
  \label{gfA}
  I_m(\tau_k)=(a_{00}^4-a_{0k}^4)^m\,\frac{\partial^m}{\partial j^m}\,\exp\left[\frac{j^2+2x_a a_{0k}^2 j}{2a_{00}^2(a_{00}^4-a_{0k}^4)} \right]_{j=0},\qquad k=1,2
\end{equation}
and
\begin{eqnarray}
  \label{gfB}
  I_{mn}(\tau_1,\tau_2)&=&(-{\rm det}\,A)^{m+n}\frac{\partial^m}{\partial j_1^m}\frac{\partial^n}{\partial j_2^n}\,\exp\left[\frac{F(j_1,j_2)}{2a_{00}^2({\rm det}\,A)^2}\right]_{j_1=j_2=0}\\
{\rm det}\,A&=&a_{00}^6+2a_{01}^2a_{02}^2a_{12}^2-a_{00}^2(a_{01}^4+a_{02}^4+a_{12}^4)\nonumber.
\end{eqnarray}
The generating function is
\begin{eqnarray}
F(j_1,j_2)&=&a_{00}^4(j_1^2+j_2^2)-2a_{00}^6(a_{01}^2j_1+a_{02}^2j_2)x_a+2 a_{00}^2(a_{12}^2j_1j_2+(a_{01}^4+a_{02}^4+a_{12}^4)(a_{01}^2j_1+a_{02}^2j_2)x_a)\nonumber\\
& &-(a_{01}^2j_1+a_{02}^2j_2)(a_{01}^2j_1+a_{02}^2j_2+4a_{01}^2a_{02}^2a_{12}^2x_a).
\end{eqnarray}
\begin{figure}
\centerline{
\setlength{\unitlength}{1cm}
\begin{picture}(12.0,9.0)
\put(0,0){\makebox(12,9){\epsfxsize=12cm \epsfbox{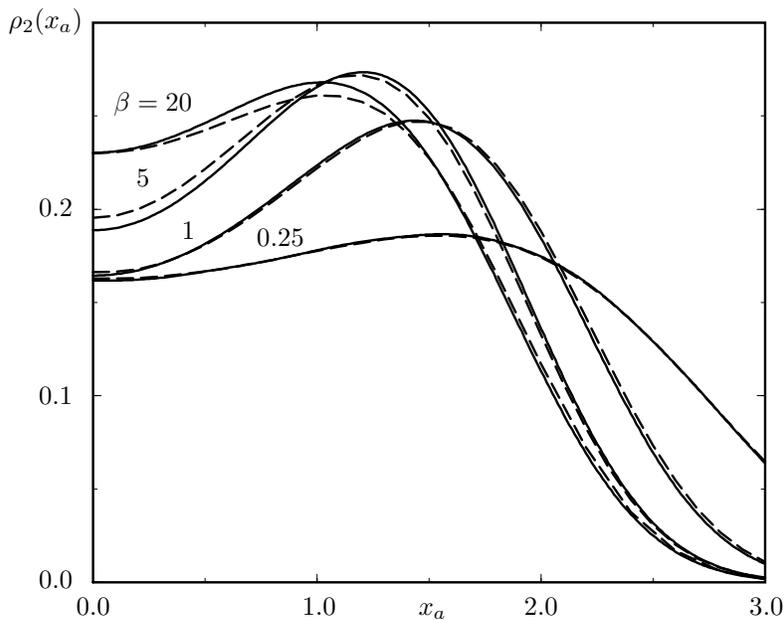}}}
\put(0.8,8.5){$\rho_2(x_a)$}
\put(6.25,0.7){$x_a$}
\put(2.2,7.4){$\beta=20$}
\put(2.5,6.4){5}
\put(3.1,5.7){1}
\put(4.1,5.6){0.25}
\put(1.7,0.7){0.0}
\put(4.7,0.7){1.0}
\put(7.65,0.7){2.0}
\put(10.65,0.7){3.0}
\put(1.2,1.05){0.0}
\put(1.2,3.5){0.1}
\put(1.2,6.00){0.2}
\end{picture}
}
\caption{\label{dwsec} Second-order particle density (dashed) compared with exact results from numerical solution of Schr\"odinger equation (solid) in a double-well at different inverse temperatures. The coupling strength is $g=0.4$.}
\end{figure}
All necessary derivatives and the imaginary time integrations in (\ref{dwcum})
have been calculated analytically. After optimizing the unnormalized second-order density (\ref{dwdens2}) in $\Omega$ we obtain the results depicted in Fig.~\ref{dwsec}. Comparing the second-order results
with the exact densities obtained from numerical solutions of the Schr\"odinger equation, we see that the deviations are strongest in the region of intermediate $\beta$, as expected. Quantum mechanical limits are reproduced very well, classical limits exactly.
\subsection{Distribution Function for the Electron in a Hydrogen Atom}
\label{csect}
With the insights gained in the last section we are prepared to apply our method to the more physical problem of an electron in a hydrogen atom,
with the attractive Coulomb interaction
\begin{equation}
  \label{coulpot}
  V({\bf r})=-\frac{e^2}{r}.
\end{equation}
Apart from the physical significance,
the
theoretical
interest in this problem originates
from
the non-polynomial bature of
the interaction. This makes
the above-developed smearing formula is essential
for finding variational perturbation expansions.
Restricting our attention to the first-order approximation for the unnormalized density, we must calculate the harmonic expectation value of the action
\begin{equation}
  \label{coulintact}
  {\cal A}_{\rm int}[{\bf r}]=\int\limits_0^{\hbar\beta}d\tau_1\,V_{\rm int}({\bf r}(\tau_1))
\end{equation}
with the interaction potential
\begin{equation}
  \label{intpot}
  V_{\rm int}({\bf r})=-\left(\frac{e^2}{r}+\frac{1}{2}{\bf r}^T\,\Omega^2\,{\bf r}\right),
\end{equation}
where the matrix $\Omega^2_{\mu\nu}$ has the form (\ref{anisoOmega}). We do not consider three more variational parameters ${\bf r}_m$ here, because this will not be relevant in a strong-coupling case like the Coulomb interaction, as we know from the last section. After optimization in $\Omega$, we compare our results for the radial distribution function
\begin{equation}
  \label{raddist}
  g({\bf r})=\sqrt{2\pi\beta}^3\,\trho({\bf r})
\end{equation}
with the precise numerical results of Storer \cite{storer}.

For the Coulomb potential,
the optimization procedure can be simplified
by setting the second optimization parameter
$x_m$ equal to zero from the outset.
This is justified by observation
made for
the double-well potential,
that the importance of  knowing
$x_m$ diminuishes for decreasing height of the central barrier.
Since the Coulomb potential has no central barrier, we may set $x_m=0$.
\subsubsection{Isotropic First-Order Approximation}
Applying the isotropic smearing formula (\ref{smeariso}) for $N=1$ to the harmonic term in (\ref{coulintact}) we easily find
\begin{equation}
  \label{isoharm}
  \meaniso{{\bf r}^2(\tau_1)}=3\frac{a^4_{00}-a^4_{01}}{a^2_{00}}+\frac{a_{01}^4}{a_{00}^4}\,r_a^2.
\end{equation}
For the Coulomb potential we obtain the local average
\begin{equation}
  \label{isocoul}
  \meaniso{\frac{e^2}{r(\tau_1)}}=\frac{e^2}{r_a}\frac{a^2_{00}}{a^2_{01}}\,{\rm erf}\left(\frac{a^2_{01}}{\sqrt{2a^2_{00}(a_{00}^4-a_{01}^4)}}r_a\right).
\end{equation}
The time integration in (\ref{coulintact}) cannot be done in an analytical manner and must be performed numerically. Alternatively we can use the expansion method introduced in Subsection~\ref{newrep} for evaluating the smearing formula in three dimensions which yields
\begin{equation}
  \label{altsmearA}
  \meaniso{{\cal A}_{\rm int}[{\bf r}]}=[\rho_0^\Omega({\bf r}_a)]^{-1}\frac{e^{-r_a^2/2 a^2_{00}}}{\pi^2 a^2_{00} r_a}\,\sum\limits_{n=0}^\infty\,\frac{H_{2n+1}(r_a/\sqrt{2 a^2_{00}})}{2^{2n+1}(2n+1)!}\,C_\beta^{(2n)} \int\limits_0^\infty dy\,y\, V_{\rm int}(\sqrt{2 a^2_{00}}\,y)e^{-y^2}H_{2n+1}(y).
\end{equation}
This can be rewritten in terms of Laguerre polynomials $L_n^\mu(r)$ as
\begin{equation}
  \label{altsmearB}
   \meaniso{{\cal A}_{\rm int}[{\bf r}]}=\sqrt{\frac{2 a^2_{00}}{\pi}}\frac{1}{r_a}\sum\limits_{n=0}^{\infty}\,\frac{(-1)^n n!}{(2n+1)!}C_\beta^{(2n)}H_{2n+1}(r_a/\sqrt{2 a^2_{00}})\int\limits_0^\infty dy\,y^{1/2}V_{\rm int}(\sqrt{2 a^2_{00}}\,y^{1/2})e^{-y}L_n^{1/2}(y)L_0^{1/2}(y).
\end{equation}
\begin{figure}
\centerline{
\setlength{\unitlength}{1cm}
\begin{picture}(12.0,9.0)
\put(0,0){\makebox(12,9){\epsfxsize=12cm \epsfbox{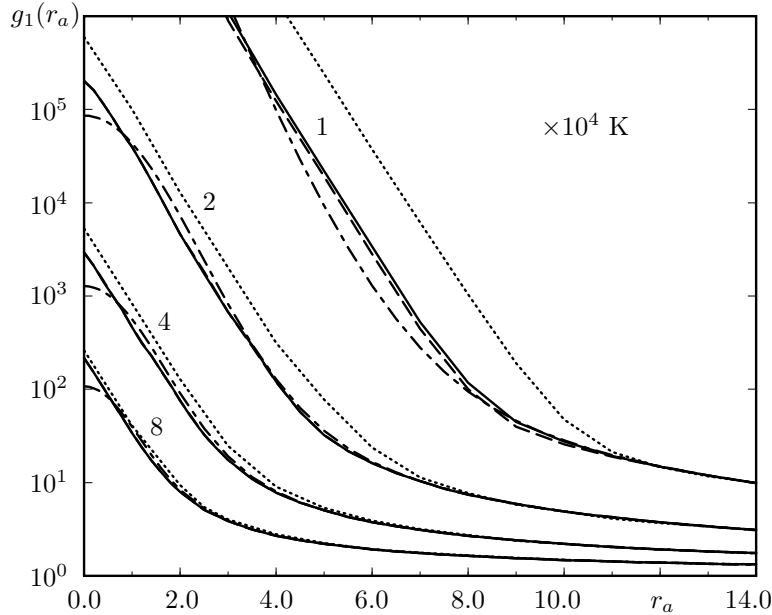}}}
\put(0.95,8.5){$g_1(r_a)$}
\put(9.45,0.7){$r_a$}
\put(8.0,7.0){$\times 10^4$ K}
\put(5.0,7.0){1}
\put(3.5,6.0){2}
\put(2.9,4.4){4}
\put(2.8,3.0){8}
\put(1.7,0.7){0.0}
\put(3.00,0.7){2.0}
\put(4.25,0.7){4.0}
\put(5.52,0.7){6.0}
\put(6.8,0.7){8.0}
\put(7.9,0.7){10.0}
\put(10.44,0.7){14.0}
\put(1.2,1.0){$10^0$}
\put(1.2,2.25){$10^1$}
\put(1.2,3.5){$10^2$}
\put(1.2,4.75){$10^3$}
\put(1.2,5.95){$10^4$}
\put(1.2,7.2){$10^5$}
\end{picture}
}
\caption[]{\label{dist} Radial distribution function for an electron--proton pair. The first-order results obtained with isotropic (dashed curves) and anisotropic (solid) variational perturbation theory are compared with Storer's numerical results \cite{storer} (dotted) and an earlier approximation derived from the variational effective potential method to first order in Ref.~\cite{kl153} (dash--dotted).}
\end{figure}
Using the integral formula \cite[Eq.~2.19.14.15]{prud}
\begin{equation}
  \label{lagint}
  \int\limits_0^\infty dx\,x^{\alpha-1}e^{-cx}L_m^\gamma(cx)L_n^\lambda(cx)=\frac{(1+\gamma)_m (\lambda-\alpha+1)_n \Gamma(\alpha)}{m! n! c^\alpha}\, _3F_2(-m,\alpha,\alpha-\lambda;\gamma+1,\alpha-\lambda-n;1),
\end{equation}
where the $(\alpha)_n$ are Pochhammer symbols, $_pF_q(a_1,\ldots,a_p;b_1,\ldots,b_q;x)$ denotes the confluent hypergeometric function, and $\Gamma(x)$ is the Gamma function, we apply the smearing formula to the interaction potential (\ref{intpot}) and find
\begin{eqnarray}
  \label{cintact}
   \meaniso{{\cal A}_{\rm int}[{\bf r}]}&=&-\frac{e^2}{\sqrt{\pi}r_a} \sum\limits_{n=0}^\infty\,\frac{(-1)^n(2n-1)!!}{2^n (2n+1)!} C_\beta^{(2n)}H_{2n+1}(r_a/\sqrt{2 a^2_{00}})\nonumber\\
& &-\frac{3}{4}\sqrt{2 a_{00}^6 \Omega^4}\frac{1}{r_a}\left\{C_\beta^{(0)}H_1(r_a/\sqrt{2 a^2_{00}})+\frac{1}{6}C_\beta^{(2)}H_3(r_a/\sqrt{2 a^2_{00}}) \right\}.
\end{eqnarray}
The first term comes from the Coulomb potential, the second from the harmonic potential. The resulting first-order isotropic form of the radial distribution function (\ref{raddist}), which can be written as
\begin{equation}
  \label{rd1}
  g_1^\Omega({\bf r}_a)=\exp[-\beta W_1^{\Omega}({\bf r}_a)]
\end{equation}
with the isotropic first-order approximation of the effective classical potential
\begin{equation}
  \label{rdveff1}
  W^\Omega_1({\bf r}_a)=\frac{3}{2\beta}{\rm ln}\,\frac{\sinh\beta\Omega}{\beta\Omega}+\frac{\Omega}{\beta}\,r_a^2\,\tanh\frac{\beta\Omega}{2}+\frac{1}{\beta}\meaniso{{\cal A}_{\rm int}[{\bf r}]},
\end{equation}
is shown in Fig.~\ref{dist} for various temperatures. The results compare well with Storer's curves \cite{storer}. Near the origin, our results are better than those obtained with an earlier approximation derived from lowest-order effective classical potential $
W_1(x_0)$ \cite{kl153}.
\subsubsection{Anisotropic First-Order Approximation}
The above results can be improved by taking care of the anisotropy of the problem. For the harmonic part of the action (\ref{coulintact}),
\begin{equation}
  \label{anisoact}
  {\cal A}_{\rm int}[{\bf r}]={\cal A}_\Omega[{\bf r}]+{\cal A}_C[{\bf r}]
\end{equation}
the smearing formula (\ref{smearaniso}) yields the expectation value
\begin{equation}
  \label{anisoharm}
  \meananiso{{\cal A}_\Omega[{\bf r}]}=-\frac{1}{2}\left\{\Omega_L^2 {a_L^2}_{00}\left(C_\beta^{(0)}+\frac{1}{2}C_{\beta,L}^{(2)}H_2(r_a/\sqrt{2 {a^2_L}_{00}})\right)+2\Omega_T^2 {a^2_T}_{00}(C_\beta^{(0)}-C_{\beta,T}^{(2)})  \right\},
\end{equation}
where the $C_{\beta,L(T)}^{(n)}$ are the polynomials (\ref{cpoly}) in which $\Omega$ is replaced by the longitudinal or transverse frequency. For the Coulomb part of action, the smearing formula (\ref{smearaniso}) leads to a double integral
\begin{equation}
  \label{anisocoul}
  \meananiso{{\cal A}_C[{\bf r}]}=-e^2\int\limits_0^{\hbar\beta}d\tau_1\,\sqrt{\frac{2}{\pi {a^2_L}_{00}(1-a_L^4)}}\int\limits_0^1 d\lambda\,\left\{1+\lambda^2\left[\frac{{a^2_T}_{00}(1-a_T^4)}{{a^2_L}_{00}(1-a_L^4)}-1\right] \right\}^{-1}\,\exp\left\{-\frac{r_a^2 a_L^4 \lambda^2}{2 {a^2_L}_{00}(1-a_L^4)} \right\}
\end{equation}
with the abbreviations
\begin{equation}
  \label{shnot}
  a^2_{L}=\frac{{a^2_{L}}_{01}}{{a^2_{L}}_{00}},\qquad a^2_{T}=\frac{{a^2_{T}}_{01}}{{a^2_{T}}_{00}}.
\end{equation}
The integrals must be done numerically and the first-order approximation of the radial distribution function can be expressed by
\begin{equation}
  \label{rd1aniso}
  g_1^{\Omega_{L,T}}({\bf r}_a)=\exp[-\beta W_1^{\Omega_{L,T}}({\bf r}_a)]
\end{equation}
with
\begin{equation}
  \label{veffaniso}
  W_1^{\Omega_{L,T}}({\bf r}_a)=\frac{1}{\beta}{\rm ln}\,\frac{\sinh\beta\Omega_L}{\beta\Omega_L}+\frac{1}{2\beta}{\rm ln}\,\frac{\sinh\beta\Omega_T}{\beta\Omega_T}+\frac{\Omega_L}{\beta}\,r_a^2\,\tanh\frac{\beta\Omega_L}{2}+\frac{1}{\beta}\meananiso{{\cal A}_{\rm int}[{\bf r}]}.
\end{equation}
This is optimized in $\Omega_L({\bf r}_a),\Omega_T({\bf r}_a)$
with the results shown in Fig.~\ref{dist}. The anisotropic approach improves the isotropic result for temperatures below $10^4$ K.
\section{Summary}
\label{discus}
We have presented variational perturbation theory for density matrices. A generalized smearing formula which accounts for the effects of quantum fluctuations was essential for the treatment of nonpolynomial interactions. We applied the theory to calculate the particle density in a double-well potential, and the electron density in a Coulomb potential, the latter as an example for nonpolynomial application. In both cases, the approximations were
satisfactory.
\acknowledgements
The work of one of us (M.B.) is supported by the Studienstiftung des
deutschen Volkes.

\end{document}